\documentclass[fleqn,usenatbib]{mnras}

\usepackage{mathptmx}

\usepackage[T1]{fontenc}

\usepackage{times}
\usepackage{color}

\usepackage{graphicx}	
\usepackage{amsmath}	
\usepackage{amssymb}	
\usepackage{hyperref}

\usepackage{subcaption}
\usepackage{longtable}
\usepackage{lscape}
\usepackage{mathtools}
\newcommand{\myaffil}[1]{$^{\rm #1}$}
\newcounter{inst}
\newcommand{\ffasource}{GLEAM\,J020507--110922}
\newcommand{\inst}[1]{\noindent%
  \refstepcounter{inst}\myaffil{\arabic{inst}\label{#1}}
}

\title[Wide-Band Spectral Variability]{Wide-band Spectral Variability of Peaked Spectrum Sources}

\author[K. Ross et al.]{K.~Ross\myaffil{\ref{ICRAR}}\thanks{E-mail: kathryn.ross@icrar.org},
N.~Hurley-Walker\myaffil{\ref{ICRAR}},
N.~Seymour\myaffil{\ref{ICRAR}},
J.~R.~Callingham\myaffil{\ref{Leiden},\ref{ASTRON}},
T.~J.~Galvin\myaffil{\ref{ICRAR}},
M.~Johnston-Hollitt\myaffil{\ref{CIC}}
\\
{\small\inst{ICRAR}\,International Centre for Radio Astronomy Research, Curtin University, Bentley, WA 6102, Australia}\\
{\small\inst{Leiden}\,Leiden Observatory, Leiden University, PO Box 9513, Leiden, 2300\,RA, The Netherlands} \\
{\small\inst{ASTRON}\,ASTRON, Netherlands Institute for Radio Astronomy, Oude Hoogeveensedijk 4, Dwingeloo, 7991\,PD, The Netherlands}\\
{\small \inst{CIC}\,Curtin Institute for Computation, Curtin University, GPO Box U1987, Perth WA 6845}\\
}

\date{Accepted 2022 March 21. Received 2022 March 18; in original form 2022 February 01}

\pubyear{2022}

\begin{document}
\label{firstpage}
\pagerange{\pageref{firstpage}--\pageref{lastpage}}
\maketitle

\begin{abstract}
Characterising spectral variability of radio sources is a technique that offers the ability to determine the astrophysics of the intervening media, source structure, emission and absorption processes. We present broadband (0.072--10\,GHz) spectral variability of 15 peaked-spectrum (PS) sources with the Australia Telescope Compact Array (ATCA) and the Murchison Widefield Array (MWA). These 15 PS sources were observed quasi-contemporaneously with ATCA and the MWA four to six times during 2020 with approximately a monthly cadence.  Variability was not detected at 1--10\,GHz frequencies but 13 of the 15 targets show significant variability with the MWA at megahertz frequencies. We conclude the majority of variability seen at megahertz frequencies is due to refractive interstellar scintillation of a compact component $\sim25$\,mas across. We also identify four PS sources that show a change in their spectral shape at megahertz frequencies. Three of these sources are consistent with a variable optical depth from an inhomogeneous free-free absorbing cloud around the source. One PS source with a variable spectral shape at megahertz frequencies is consistent with an ejection travelling along the jet. We present spectral variability as a method for determining the physical origins of observed variability and for providing further evidence to support absorption models for PS sources where spectral modelling alone is insufficient. 

\end{abstract}




\section{Introduction}
\label{sec:intro}
Variability at radio wavelengths of active galactic nuclei (AGN) has the potential to reveal their radio structures, astrophysical properties, and the medium between the observer and the source. Long-duration variability at radio wavelengths has previously been shown to provide insight into a range of intrinsic phenomena including young jets \citep{2020ApJ...896...18P, 2020ApJ...905...74N}, jet interactions or shocks \citep{Jamil_2010_iShocks}, flare events and adiabatic expansion \citep{2008A&A...485...51H}, oscillating jet orientation \citep{2011A&A...526A..51K}, or the nature of a surrounding ionized medium \citep{tingay2015spectral,2018MNRAS.475.3493B}. 

Short duration (hours to days) variability in the gigahertz regime is largely attributed to extrinsic propagation effects such as interstellar scintillation \citep[ISS;][]{2008ApJ...689..108L,2018MNRAS.474.4396K}. Characterising the timescales and size of modulation due to ISS can provide information on source morphologies on micro-arcsecond ($\mu$as) scales \citep{narayan1992physics, 1998MNRAS.294..307W}. Furthermore, variability at low frequencies ($<1$\,GHz) has also been attributed to ISS, particularly refractive ISS  \citep[RISS;][]{hunstead_1972,Rickett1986RISS,mwats,hancock2019refractive}. 

Previously, \citet[][hereafter  R21]{2021MNRAS.501.6139R}, conducted one of the largest searches for spectral variability at radio frequencies to date. R21 surveyed over 21,000 sources with the Murchison Widefield Array \citep[MWA;][]{tingay2013murchison} over 100--231\,MHz with a two epochs separated by roughly one year. R21 introduced the variability index parameter (VIP) to detect variability across a wide spectral range, and the measure of spectral shape (MOSS) parameter to classify the type of variability. R21 found a range of spectral variability, from uniform increases in flux density across the observing band to various changes in spectral shape. Furthermore, R21 also found that AGN with a peak in their spectral energy distributions (SEDs) appear to be more variable than typical power-law AGN. These peaked-spectrum (PS) sources are typically also compact ($\leq$20\,kpc); see \citet{2021A&ARv..29....3O} for a comprehensive review. PS sources have been shown to have a higher scintillation index for interplanetary scintillation (IPS) with the MWA \citep{ipsII}. Likewise, high-resolution imaging with VLBI found that sources with compact morphologies also had high scintillation indices \citep{2021MNRAS.tmp.2744J}. PS sources have also been shown to vary significantly on decade-long timescales attributed to renewed AGN activity and young, evolving jets \citep{2017FrASS...4...38W,2020ApJ...905...74N}. 

The cause of the low-frequency absorption producing the spectral peak of PS sources is still largely debated between two competing theories: synchrotron self-absorption (SSA) or free-free absorption (FFA). The first case, often considered the `youth' scenario \citep{o1997constraints}, suggests their compact size is likely due to the jets being young and having formed within the past $\sim$10$^5$\,years \citep{1998A&A...337...69O} and that SSA occurs due to high brightness temperatures at low frequencies. Alternatively, the FFA case, often referred to as the `frustration' scenario \citep{1984AJ.....89....5V}, suggests the radio jet/lobe is prevented from growing due to a surrounding cloud of dense ionized plasma \citep{1997ApJ...485..112B}. Unfortunately, distinguishing between these two scenarios requires large spectral coverage in the optically-thick regime (below the spectral turnover), and complex (often inconclusive) spectral modelling \citep{2003AJ....126..723T,2015ApJ...809..168C}. 
Furthermore, previous variability monitoring of PS sources found many displayed a temporary peak in their SED and lost their PS source classification \citep[][R21]{torniainen2005long}. Such temporary PS sources were considered likely to be blazars (i.e. a radio source with one jet pointed towards the observer), rather than compact symmetric objects \citep{1996ApJ...463...95T}. 

One of the key issues in radio variability, at all frequencies, is distinguishing between intrinsic and extrinsic origins \citep{2008ApJ...689..108L}. Once the origin of variability is determined, it can be used to inform the physical properties of the source itself. Given recent findings that PS sources appear to be a more variable population compared to typical AGN \citep[][R21]{ipsII}, this population provides a unique opportunity to study variability mechanisms. Furthermore, the variability above and below the spectral peak (in the optically thin and optically thick regimes respectively) may be due to separate physical mechanisms. \citet{tingay2015spectral} monitored the nearby PS source, PKS\,B1718--649, with the ATCA for almost two\,years, with a large spectral coverage of 1--10\,GHz. The vast spectral coverage was able to cover the optically thick and optically thin regimes as well as the spectral turnover at $\sim3$\,GHz, which allowed for confident spectral modelling of both SSA and FFA spectral models. \citet{tingay2015spectral} also detected variability across the entire sampled spectrum of PKS\,B1718--649. By combining low- and high-frequency observations to search for spectral variability, \citeauthor{tingay2015spectral} were able to refine the causes of variability below and above the spectral turnover as being due to different physical processes. Furthermore, low-frequency variability was found to be caused by changes in the free-free optical depth, as the magnitude of variability across the spectrum was inconsistent with SSA. While the spectral modelling provided tentative evidence for an FFA spectral model over an SSA spectral model, the cause of the low-frequency variability being due to variations in the free-free optical depth added further evidence in support of a FFA spectral model.

Spectral variability, therefore, has the potential to distinguish between intrinsic and extrinsic origins of variability. Until recently, surveys of spectral variability have been limited by single/small sample sizes \citep{tingay2015spectral}, narrow spectral coverage \citep{hunstead_1972,Fanti_1979,mwats}, only gigahertz frequency coverage \citep{2020ApJ...905...74N,2021AN....342.1212W} or combining non-simultaneous spectral coverage \citep{torniainen2005long}. Furthermore, the low-frequency spectral variability (in the optically thick regime of PS sources) appears to have distinct properties and origins compared to variability at gigahertz frequencies. With the development of the MWA, and leading into the next generation of telescopes such as the Square Kilometre Array low frequency array (SKA\_LOW), surveys of large spatial regions/population sizes with significant temporal and spectral coverage are now becoming achievable. 

In this paper, we build on the work of R21 to study the spectral variability (0.07--10\,GHz) of 15~PS sources to determine the origins of their variability and absorption at megahertz frequencies. The combined simultaneous observations from the MWA (0.07-0.23\,GHz) and the Australia Telescope Compact Array (ATCA, 1-10.GHz) over one year make this survey a unique study of broad spectral variability. Section~\ref{sec:obs_srces} outlines the selection process of the 15~PS sources in this study. The observational and data reduction strategies for the MWA and the ATCA are described in Section~\ref{sec:obs_mwa} and Section~\ref{sec:obs_atca} respectively. The spectral models and fitting routines are described in Section~\ref{sec:spectral_modelling}. The results are summarised in Section~\ref{sec:results} and detailed analysis and discussion of individual sources is presented in Section~\ref{sec:discussion}. All coordinates are in J2000. 

\section{Source Selection}
\label{sec:obs_srces}
The main goal of this variability monitoring campaign was large quasi-contemporaneous spectral coverage using the ATCA and the MWA for a small number of targets. 

The sample of sources were selected for follow up monitoring according to several criteria: 
\begin{enumerate}
    \item classified as a PS source by \citet{Callingham_2017};
    \item predicted flux density $\geq10$\,mJy at 9\,GHz;
    \item observed to show spectral variability in R21 with a variability index parameter (VIP) $\geq58.3$ according to Equation~\ref{eq:VIP}.
\end{enumerate}

We selected PS source targets based on criteria (i) for a reliable PS source classification. \citeauthor{Callingham_2017} combined flux density measurements from the GaLactic and Extragalactic All-Sky MWA \citep[GLEAM;][]{wayth2015gleam} ExGal data release \citep{2017MNRAS.464.1146H} with flux density measurements from either the Sydney University Molonglo Sky Survey \citep[SUMSS;][]{2003MNRAS.342.1117M} or the NRAO VLA Sky Survey \citep[NVSS;][]{condon+98} to identify PS sources with a spectral peak between 72\,MHz and 1.4\,GHz. Sources were classified as a PS source if they either showed a spectral peak or curvature within the GLEAM band (72 -- 231\,MHZ), or a power-law spectrum with a positive spectral index. As the frequencies of spectral peaks of the PS sources presented by \citeauthor{Callingham_2017} are below 1.4\,GHz, our monitoring with the MWA and the ATCA with a spectral coverage of 0.072--10\,GHz covers both the optically thin and optically thick regimes of each of our targets. 

For criterion (ii), we calculated the spectral index of a power-law spectral model and predicted the flux densities at 9\,GHz. The power-law was fit using the GLEAM flux density measurement at 220\,MHz and the flux density at either 843\,MHz or 1.4\,GHz, based on the availability of either SUMSS or NVSS. This criterion ensures we have enough signal-to-noise in the ATCA data to probe variability at $<10\%$ level.

We selected the 15 most promising targets that satisfied all three criteria as the sources for this study. Initial results of variability for R21 identified 15~targets that satisfied all three criteria. Criterion (iii) used the variability index parameter (VIP) according to: 

\begin{equation}
    \mathrm{VIP} = \sum^{n}_{i=1} \frac{\big(S_{1}(i)- S_{2}(i)\big)^2}{\sigma_i^2},
    \label{eq:VIP}
\end{equation}
where $S_1(i)$ and $S_2(i)$ are the flux densities in the first and second epoch in a given sub-band $i$, respectively, and $\sigma_i$ is the combined uncertainty of each flux density added in quadrature. The VIP is a measure of how many flux density measurements in the second epoch differ from those in the first epoch, and by how much. As part of our initial results for R21, we identified 15~targets that satisfied criteria (i) and (ii) and had a VIP that implied they were variable. However, five of our proposed PS targets were later excluded from the final catalogue of variable sources in R21 due to lower mosaic quality and higher median VIP in those regions. Despite the lower mosaic quality, these five targets were included in this variability study as their measured VIP was at least two times greater than the significance cut off used for the other areas of the mosaic and significantly larger than the median value in the poor quality regions. As such, the variability of these five targets were considered more significant than the variability due to the poorer mosaic quality. The measure of spectral shape (MOSS) parameter for these targets was also calculated according to: 
\begin{equation}
    \mathrm{MOSS} = \sum^{n}_{i=1} \frac{(\widetilde{\mathrm{diff}}- \mathrm{diff}(i))^2}{\sigma_{i}^2}
\end{equation}
where $\widetilde{\mathrm{diff}}$ is the median of the differences between the flux density over all frequencies, $\mathrm{diff}(i)$ is the difference of the flux densities between the two epochs at frequency $i$, and $\sigma_{i}$ is the combined uncertainty of each flux density added in quadrature.
The calculated VIP and MOSS for all 15 PS targets are presented in Table~\ref{tab:targets}.

Only 10 of our original 15~targets were classified as variable in R21 but all have a VIP$>$58.3. Furthermore, all 15 satisfied criteria (i) and (ii). All 15~targets were included in this study. A summary of the 15~targets monitored in this study and the observations used in the analysis can be found in Table~\ref{tab:targets}.

\begin{table}
\centering
\begin{tabular}{|l|l|l|l|l}
\hline
GLEAM Name & $S_\mathrm{151MHz}$ (Jy) & VIP & MOSS  \\
\hline
    J001513-472706 & 0.50 & 463 & 35 \\ %
    J015445-232950 & 0.65 & 331* & 11* \\ %
    J020507-110922 & 1.36 & 1092 & 48 \\ %
    J021246-305454 & 0.29 & 125* & 11*  \\ %
    J022744-062106 & 0.48 & 431 & 8  \\ %
    J024838-321336 & 0.41 & 264 & 38  \\ %
    J032213-462646 & 0.42 & 336 & 26  \\ %
    J032836-202138 & 0.55 & 290 & 25  \\ %
    J033023-074052 & 0.33 & 816 & 46  \\ %
    J042502-245129 & 0.63 & 431* & 76*  \\ %
    J044033-422918 & 1.86 & 1095 & 15  \\ %
    J044737-220335 & 2.67 & 767* & 104*  \\ %
    J052824-331104 & 0.64 & 173* & 38*  \\ %
    J223933-451414 & 1.43 & 2796 & 95  \\ %
    J224408-202719 & 0.39 & 226 & 19  \\ %
\hline
\end{tabular}%
\caption{Targets chosen for monitoring. $S_\mathrm{151MHz}$ is as reported in the GLEAM catalogue \citep{2017MNRAS.464.1146H}. All sources are compact within GLEAM, thus the GLEAM names are also accurate coordinates to $\sim2'$. We present the corresponding variability index parameter (VIP) and measure of spectral shape (MOSS) parameter. A VIP$\geq58.3$ was classified as variable and a MOSS$\geq36.7$ was classified as changing spectral shape, according to R21. Five targets did not meet criteria (iii) as they were cut from the final catalogue of variable sources presented by R21, as discussed in Section~\ref{sec:obs_srces}. The presented VIP and MOSS values for these targets are denoted with a $*$ and has been calculated as part of this work. } 
\label{tab:targets}
\end{table}


\section{Observations}
Each target was observed six separate times during 2020. However, due to some observational difficulties, discussed in detail in Sections~\ref{sec:obs_mwa} and \ref{sec:obs_atca}, some epochs were discarded from analysis for both the ATCA and the MWA. Table~\ref{tab:epochs} summarises the telescope configurations and observation information for each epoch, we also include the two original GLEAM epochs (MWA Phase I from R21). 

\begin{table*}
\centering
\begin{tabular}{|l|l|l}
\hline
Epoch & Notes on MWA Observations     & Notes on contemporaneous ATCA Observations  \\
\hline
    August to September 2013 & GLEAM Year 1, MWA Phase I & No Data \\
    August to December 2014 & GLEAM Year 2, MWA Phase I & No Data \\
    January~2020 & No Data & 6A configuration, 2.1, 5.5 and 9\,GHz observed  \\
    March~2020 & No Data &  6D configuration, incorrect central frequency for X-band, \\ & & 9\,GHz data omitted for \\
    & & GLEAM\,J001513--472706, \\ & & GLEAM\,J223933--451414 and \\ & & GLEAM\,J224408--202719\\
    April~2020 & Extended Phase II configuration, & 6A configuration, 2.1, 5.5 and 9\,GHz observed \\
    & Daytime observations, omitted for \\
    & GLEAM\,J020507--110922, GLEAM\,J024838--321336 & \\
    May~2020 & Extended Phase II configuration & 6A configuration, 2.1, 5.5 and 9\,GHz observed \\
    July~2020 & Extended Phase II configuration & No Data \\
    September~2020 & Extended Phase II configuration & No Data    \\
    October~2021 & No Data & H168 configuration, 5.5 and 9\,GHz observed for \\
    & & GLEAM\,J001513--472706 and GLEAM\,J020507--110922 \\
\hline
\end{tabular}%
\caption{The dates of each epoch used for the variability analysis, details of the telescope configuration for both MWA and the ATCA and any relevant notes on the observations or data omitted.} 
\label{tab:epochs}
\end{table*}

\subsection{MWA}
\label{sec:obs_mwa}

The MWA observations were scheduled to match the awarded ATCA observations\footnote{The project code for MWA observations is G0067, PI: Ross.}. Unfortunately, the MWA~Phase~\textsc{II} was in the compact configuration during the January and March~2020 ATCA observations\footnote{The side-lobe confusion of the MWA in the compact configuration is often large enough that the scientific use of the final image is limited. For details of the configurations of MWA~Phase~\textsc{II}, see \citep{2018PASA...35...33W}}. As such, these observations were omitted from our analysis. All subsequent MWA observations were obtained with the MWA in the extended Phase~\textsc{II} configuration. Two further epochs were observed with the MWA in July and September~2020 without contemporaneous ATCA observations. Thus there are a total of four usable MWA epochs over six months of 2020 with two taken within 48 hours of the ATCA observations in April and May~2020. Furthermore, the GLEAM South Galactic Pole observations from 2013 and 2014 \citep{2021PASA...38...14F}, used by R21, were also considered to make a roughly six year time baseline. 

The observational strategy for the MWA relied on targeted two-minute snapshots with the target source approximately centred in the primary beam. Due to the large field of view of the MWA, the sensitivity is fairly consistent within $\sim$5\,degrees of the pointing centre. These targeted snapshots were taken at five different frequency bands of 30.72\,MHz bandwidth and centred at 87.7, 118.4, 154.2, 185.0, and 215.7\,MHz to match the frequencies used in GLEAM survey \citep{wayth2015gleam}. High elevations were required for good sensitivity, so the April and May~2020 observations were taken during the day. Where possible, the Sun was placed in a null of the primary beam to reduce its effect on the observations. In the April observations, each target had three snapshots for each frequency band; for subsequent epochs, each target had six target snapshots. 

We employed a similar strategy as used for the GLEAM-X survey data reduction\footnote{The GLEAM-X pipeline can be found and downloaded here: \url{https://github.com/tjgalvin/GLEAM-X-pipeline}}. No calibration scans were taken, instead the latest sky model, GLEAM Global Sky Model (GGSM), was used to calculate calibration solutions for each snapshot (Hurley-Walker et al. submitted). For the region containing all 15 of our targets, this model is largely derived from GLEAM ExGal \citep{wayth2015gleam,2017MNRAS.464.1146H}. Following the same reduction strategy as GLEAM-X, any known bad tiles were flagged before initial calibration solutions were calculated with respect to the GGSM. These solutions were inspected and any further bad tiles were flagged before applying the solutions. If the calibration was unable to converge on solutions, solutions from an observation taken around a similar time with a similar pointing were applied. 

Initial images were made using \textsc{wsclean} \citep{2014MNRAS.444..606O} with a Briggs weighting of robust weighting of +0.5 \citep{2021PASA...38...53D}. Images were visually inspected to ensure calibration was appropriate and assess the effects of the Sun for any day-time observations or bright sources known to reduce image quality\footnote{A list of these sources can be found in  Table~2 of \citet{2017MNRAS.464.1146H}.}. In the April observations, despite placing the Sun in a null of the primary beam where possible, due to the frequency dependence of the primary beam, some pointings resulted with the Sun within the images. This significantly increased the noise in the images and large scale artefacts across the entire image. 

For GLEAM\,J020507--110922 and GLEAM\,J024838--321336, the location of the Sun resulted in at least twice the local root-mean-squared (RMS) noise in snapshot images for each frequency compared to other targets. As a result, this epoch for these targets was emitted. For remaining day-time observations, imaging parameters were adjusted to reduce the power of the Sun. Since the Sun is resolved and this study was only interested in unresolved (with the MWA) bright sources, short baselines of the MWA were removed when producing images that contained the Sun in the primary beam. The short baselines were tapered out using the \texttt{minuv-l} and \texttt{taper-inner-tukey} options of \textsc{wsclean} to create a gradual taper of shorter baselines rather than a sharp cut. Both the \texttt{minuv-l} and \texttt{taper-inner-tukey} were set to 50\,$\lambda$ to cut baselines less than 50\,$\lambda$ and minimise baselines between 50 and 100\,$\lambda$. The $uv$-taper also reduced the effect of bright, resolved galaxies, like Fornax-A, which were also occasionally in the field of view.  

Once satisfactory images were produced, the flux density scale and position shifting were corrected, to account for miscalculations of the primary beam and effects from the ionosphere, respectively. A correction was derived and applied using \textsc{flux\_warp} \citep{2020PASA...37...37D} and \textsc{fits\_warp} \citep{2018A&C....25...94H}, both of which use a subset of the Global GLEAM Sky Model (GGSM). Only unresolved (according to GLEAM), bright (signal to noise ratio $\geq10$) and isolated (to within 5\,arcminutes) sources were considered in the reference catalogue to ensure a reliable model of the flux density scale and compared to the GGSM catalogue (Hurley-Walker et al. submitted). Corrected images were stacked together to create a small mosaic of 5,000 by 5,000 pixels with the target at the centre. Images were stacked using \textsc{swarp} \citep{2002ASPC..281..228B} and coadded with inverse-variance weighting using the RMS noise of the images. Due to the large field of view of the MWA, some observations covered multiple targets. To decrease the overall RMS of stacked images, any observations where the target was within the field of view of the MWA were included in the stacking, even if it was not a targeted observation.   

The variable ionospheric conditions during the observations can result in a residual blurring effect. To correct for this, a blur correction was applied to the resampled, coadded mosaics by generating point-spread-function (PSF) maps. Firstly, the background and noise maps of the mosaics were generated using the Background and Noise Estimation tool (\textsc{BANE}). An initial shallow source finding was run on the resultant mosaics using \textsc{Aegean}\footnote{\url{https://github.com/PaulHancock/Aegean}} \citep{2012MNRAS.422.1812H,2018PASA...35...11H}. For this shallow source finding, the ``seed'' clip was set to 10, i.e. only pixels with a flux density 10 times the local RMS noise were used as initial source positions. The output catalogue from the shallow run of \textsc{Aegean} was cut to only include unresolved sources. This catalogue was then used to produce a measured PSF map for the mosaic. The measured PSF map was used as input for a further run of \textsc{Aegean} to account for the variable PSF of the mosaic. The generated catalogue of sources from the second run of \textsc{Aegean} was used to generate a new PSF map with the right blur correction, which we applied to the mosaic to correct for the ionosphere. Resolved sources were excluded from the catalogue for this blur correction. 

A final correction for any large scale flux density variations across the blur corrected mosaic was applied using \textsc{flux\_warp} again. This correction was of the order of 2\%--10\% depending on the frequency and whether the observations were taken during the day. As with the first run of \textsc{flux\_warp}, a reference catalogue of bright, unresolved and isolated sources was used to ensure a reliable model of the flux density scale and compared to the GGSM catalogue. The GGSM catalogue was used as a prior catalogue for source positions for \textsc{Aegean}'s priorised fitting. Furthermore, any sources that were previously classified as variable by R21 were excluded from the reference catalogue.

A final source-finding of the blur and flux density scale corrected mosaics using \textsc{BANE} and \textsc{Aegean} produced the catalogue used in variability analysis. 

\subsection{ATCA}
\label{sec:obs_atca}

In 2020, four observations of the 15~targets were taken in January, March, April and May. Observations were taken at L-band (central frequency 2.1\,GHz), and C/X-band (central frequencies 5.5\,GHz and 9.5\,GHz). The bandwidth in all cases was 2\,GHz \citep{2011MNRAS.416..832W}. For the January and March epochs, the observing strategy was two 12-hour blocks on consecutive days, each of which was devoted to a single ATCA band. The April and May epochs each had an 18\,hour observing block, and frequency switching was used between the two bands. In all epochs, two-minute snapshots were taken of the target sources sandwiched between secondary phase calibrator observations. Secondary calibrators were shared between targets when both targets had an angular separation less than 10\,degrees to the secondary calibrator in order to reduce slew overheads. The $(u,v)$-coverage was more complete in the April and May epochs compared to the January and March epochs as there was a larger time gap between snapshots due to the frequency switching. All epochs were observed in a 6-km array configuration; for specific array configurations in each epoch, see Table~\ref{tab:epochs}. 

The same primary bandpass calibrator, PKS\,B1934--638, was observed for each epoch and used for estimates of the overall instrumental errors. Furthermore, we use the measured flux density of PKS\,B1934-638 in each epoch to compare with our sources to assess the variability of the target sources.

Due to an error in scheduling, GLEAM\,J001513--472706, GLEAM\,J223933--451414 and GLEAM\,J224408--202719 were observed at 9\,GHz in March while PKS\,B1936--638 was observed at 9.5\,GHz. As a result, the 9\,GHz observations for the sources in March were discarded to avoid applying an inaccurate calibration solution and inducing artificial variability. 

The majority of data reduction was completed using \textsc{casa 6.4} \citep{2007ASPC..376..127M} after first converting the data into a measurement set via the \textsc{miriad} \citep{1995ASPC...77..433S} task \texttt{atlod}. Data were processed using the same reduction procedure\footnote{The code used to process all the ATCA data can be found here: \url{https://github.com/astrokatross/ATCA_datareduction}}, which we briefly describe here. After initial flagging for radio frequency interference (RFI), observations were split into a separate measurement set with the primary bandpass calibrator and associated secondary calibrator. An initial round of bandpass and gain calibration solutions were calculated using just the primary calibrator. Then a second round of gain calibration solutions were calculated using the primary and secondary calibrators. The flux density scale was estimated using PKS\,B1934--638 as a flux density standard \citep{reynolds1994revised}. Further RFI flagging was performed on the calibrated measurement set before an initial model image was created using the interactive \texttt{tclean} on a Multi-Frequency Synthesis (MFS) image of the entire bandwidth. Three rounds of phase self-calibration were performed on the target source using the created model image. For all targets, in the 5.5\,GHz and 9.5\,GHz bands, no other sources sources were detected in the field of view. However, the larger field-of-view for the 2.1\,GHz band resulted in an occasional nearby source in the image field. Where appropriate, these other sources were included in the model used for self-calibration. Self-calibration solutions were calculated by combining the entire ATCA band to increase signal to noise, and applied without flagging any sections that were unable to converge on a solution. Targets were split into smaller spectral windows for imaging (to create the model) and flux density measurements.
Observations at 2.1\,GHz were split into eight spectral windows, 5.5\,GHz into five and 9.5\,GHz into four. Such binning ensured roughly equal fractional bandwidth per spectral band. The flux density for each spectral band was measured using the \texttt{uvmodelfit} function in \textsc{casa}. A rough initial source position was given based on the MFS image but allowed to vary.  

The flux densities of secondary calibrators and the primary bandpass calibrator were also measured using \texttt{uvmodelfit}. These measurements were used to estimate systematic errors on the flux density measurements of targets and assess the significance of any variability.

In October~2021, opportunistic follow-up observations observations with the ATCA at 5.5\,GHz and 9\,GHz during Director's Time were undertaken of GLEAM\,J001513--472706 and GLEAM\,J020507--110922 in the H168 configuration. The observational strategy differed slightly to the 2020 monitoring. Targets were observed with 10\,minute scans over several hours. 

\section{Spectral Modelling}
\label{sec:spectral_modelling}

We fit spectral models to each source at each epoch to determine the underlying absorption mechanism. There are two main mechanisms for absorption at low frequencies: synchrotron self-absorption (SSA) or free-free absorption (FFA). The SSA model assumes the electron energy distribution for a single homogeneous synchrotron emitting region is described by a non-thermal power-law with index $\beta$. A spectral turnover occurs in a SSA model when the photons from the source are scattered by the relativistic electrons in the plasma. The low-energy photons are more likely to be scattered repeatedly resulting in them appearing to be "re-absorbed" by the plasma. The SSA model can be described according to Equation~\ref{eq:singSSA} of \citet{1966ApJ...146..621K}, where $\nu_p$ is the frequency where the source becomes optically thick (i.e. the optical depth, $\tau_\nu$, is unity). Namely,

\begin{equation}
    \begin{aligned}
        S_\nu &= S_\mathrm{norm}\left(\frac{\nu}{\nu_{p}}\right)^{\frac{\beta-1}{2}}\left[\frac{1-e^{-\tau_{\nu}}}{\tau_{\nu}} \right], \\
        \mathrm{where} \\
        \tau_\nu &= \left( \frac{\nu}{\nu_p} \right)^{\frac{-(\beta+4)}{2}}.
        \label{eq:singSSA}
    \end{aligned}
\end{equation}

Alternatively, the FFA model assumes a process of inverse bremsstrahlung or free-free absorption, where an ionized plasma screen is causing the absorption of the photons emitted by the relativistic electrons from the source \citep{1997ApJ...485..112B,2003AJ....126..723T,2015ApJ...809..168C}. In this scenario, the electrons emit photons described by a non-thermal power-law distribution, using $\alpha$ as the spectral index of the synchrotron emission, where $\alpha=\left( \beta -1 \right)/2$ for the electron energy distribution as described by Equation~\ref{eq:singSSA}. Several variations of FFA models exist that account for variations in screen morphology (either homogeneous or inhomogeneous) and whether the absorption is external or internal to the emitting electrons. In this work, we only consider FFA models with an external ionized screen that is either homogeneous or inhomogeneous since internal free-free absorption has been shown to poorly replicate observed spectra of PS sources \citep[e.g.][]{2015ApJ...809..168C}. 

The external homogeneous FFA model assumes a uniform ionized absorbing screen covers the entire emitting source. For a screen with optical depth $\tau_\nu$, the external homogeneous FFA model is written \citep{1997ApJ...485..112B}:

\begin{equation}
    \begin{aligned}
        S_\nu &= S_\mathrm{norm}\nu^{\alpha}e^{-\tau_\nu}, \\
        \mathrm{where} \\
        \tau_\nu &= \left(\frac{\nu}{\nu_p}\right)^{-2.1}
        \label{eq:FFA}
    \end{aligned}
\end{equation}
where $\nu_p$ is the frequency where the free-free optical depth equals unity. 

The inhomogeneous FFA model is an external FFA model where the absorbing ionized cloud has a range of optical depths. The inhomogeneous FFA model was first presented by \citet{1997ApJ...485..112B}, who modelled the interaction of the radio jets with the surrounding interstellar medium (ISM). \citet{1997ApJ...485..112B} proposed the jets create shocks in the ISM as they propagate from the AGN, producing regions of shocked gas with spatially variable optical thickness. To derive the spectral model of such a scenario, \citet{1997ApJ...485..112B} assumed the range of optical depths can be described by a power-law distribution with index $p$ according to:
\begin{equation}
    \tau_{\mathrm{ff}} \propto \int (n_e^2 T_e^{-1.35})^p dl,
    \label{eq:inFFA_opticaldepths}
\end{equation}
\noindent where $n_e$ is the free electron density and $T_e$ is the electron temperature. We assume $p>-1$, otherwise as this model reduces to the homogeneous condition. 
By assuming the scale of the lobes is much larger than the scales of the inhomgeneities in the ISM and the shocks, \citet{1997ApJ...485..112B} represent the inhomogeneous FFA model as:

\begin{equation}
    S_\nu = S_\mathrm{norm}(p+1)\gamma\left[p+1, \left( \frac{\nu}{\nu_p} \right)^{-2.1}  \right]\left( \frac{\nu}{\nu_p} \right)^{2.1(p+1) +\alpha}
    \label{eq:inFFA}
\end{equation}
\noindent
where $\gamma$ is the lower incomplete gamma function of order $p+1$. In this model, the spectral index of the optically thick regime is described by $\alpha_\mathrm{thick} = \alpha - 2.1(p+1)$ \citep{1997ApJ...485..112B}.

Each of the SSA, FFA, and inhomogeneous FFA models assume a non-thermal synchrotron emission power-law distribution of the relativistic electrons. However, a continuous injection model \citep{1962SvA.....6..317K} predicts that the higher-energy electrons cool more quickly than the lower-energy electrons, presenting as a spectral steepening at frequencies higher than a break frequency, $\nu_\mathrm{break}$. We introduce an exponential multiplicative factor, $e^{-\nu/\nu_\mathrm{break}}$, into the SSA, homogeneous FFA and inhomogeneous FFA models to represent the spectral steepening \citep{2015ApJ...809..168C}. We therefore fit a total of six spectral models: SSA, SSA with an exponential break, external homogeneous FFA, external homogeneous FFA with a spectral break, external inhomogeneous FFA and an external inhomogeneous FFA with a spectral break.

We fitted each spectral model using the \verb|UltraNest| package\footnote{\url{https://johannesbuchner.github.io/UltraNest/}} \citep{2021JOSS....6.3001B}. \verb|UltraNest| uses a nested sampling Monte Carlo algorithm MLFriends \citep{2017arXiv170704476B,2016S&C....26..383B} to derive the Bayesian evidence and posterior probability distributions. We assumed a Gaussian distribution for the likelihood of each parameter and used a reactive nested sampler. As discussed in Section~\ref{sec:results}, we detected no significant variability with the ATCA across the 2, 5.5 or 9\,GHz frequency sub-bands. As a result, ATCA flux densities were combined over time per sub-band to create an average flux density with 17 unique spectral points per source. This average ATCA spectrum was used to fit each MWA epoch over using individual ATCA epochs. 


To compare spectral models, we calculate the Bayes factor, $K$: 
\begin{equation}
    K = e^{\log{z_1}-\log{z_2}}
    \label{eq:Bayesfactor}
\end{equation}
\noindent
for each pair of models where $z_i$ is the maximum likelihood of the model $i$. Models with fewer parameters have a higher likelihood, thus the Bayes factor is robust against preferring over-fitting. Assuming the physical mechanism causing the absorption in the SED is constant between epochs, we can determine the most likely spectral model based on all epochs. We calculate the average log likelihood of each model per source and conclude the most likely model is that with the largest average log likelihood. We calculate the Bayes factor, according to Equation~\ref{eq:Bayesfactor}, for the preferred model to the second most likely model to determine the significance of the likelihood. 
If $K \geq 100$, the likelihood of the first model is strongly more likely. If $ K < 10$, the first model is more likely but there is less evidence of support. We present the average log likelihood (averaged over all the epochs), $\log(i)$ for each model in Table~\ref{tab:targets_specmodels}.

\begin{table*}
\centering
\begin{tabular}{l|l|l|l|l|l|l|l|l}
\hline
GLEAM Name & $\log{L_{\mathrm{SSA}}}$ & $\log{L_{\mathrm{FFA}}}$ & $\log{L_{\mathrm{inFFA}}}$ & $\log{L_{\mathrm{SSAb}}}$ & $\log{L_{\mathrm{FFAb}}}$ & $\log{L_{\mathrm{inFFAb}}}$ & Best Model & Variability \\
\hline
    J001513-472706 & 75 & 93.2 & 121.4 & 115.3 & 117.6 & 125.2 & inFFA(b) & Brightness change at $\leq$231\,MHz \\
    J015445-232950 & 30.5 &  61 & 107.6 &  98.5 & 105.9 & 120.6 & inFFAb & Variable spectral shape at $\leq$231\,MHz  \\
    J020507-110922 & 19.0 & 47.7 & 84.6 & 74.6 & 82.6 & 99.9 & inFFAb & Variable spectral shape at $\leq$231\,MHz \\
    J021246-305454 & 113.8 & 40.9 & 148.2 & 154.3 & 152.9 & 156.0 & inFFAb* & Brightness change at $\leq$231\,MHz  \\
    J022744-062106 & -38.4 & -20.4 & 98.7 & 76.2 & 77.1 & 98.0 & inFFA(b) & No variability detected \\
    J024838-321336 & 65.0 &  76.6 & 87.5 & 119.8 & 116.3 & 115.3 & (SSA)b & Variable spectral shape at $\leq$231\,MHz  \\
    J032213-462646 & -167.9 & -256.4 & 83.8 & 56.0 & 44.0 & 89.0 & inFFA(b) & Brightness change at $\leq$231\,MHz  \\
    J032836-202138 & 12.3 & 62.9 & 131.4 & 124.2 & 136.3 & 151.7 & inFFAb & Brightness change at $\leq$231\,MHz \\
    J033023-074052 & 25.6 & 7.9 & 104.2 & 106.6 & 101.5 & 114.4 & inFFAb & Brightness change at $\leq$231\,MHz  \\
    J042502-245129 & -291.7 & -287.3 & 76.9 & 49.7 & 27.2 & 102.4 & inFFAb & Brightness change at $\leq$231\,MHz  \\
    J044033-422918 & -291.9 & -242.3 & 51.7 & -73.7 & -76.8 & 82.4 & inFFAb & Brightness change at $\leq$231\,MHz \\
    J044737-220335 & -1816.0 & -2351.6 & 29.8 & -876.2 & -1210.2 & 76.2 & inFFAb & Brightness change at $\leq$231\,MHz  \\
    J052824-331104 & -71.8 & -78.7 & 90.9 & 74.2 & 50.5 & 104.7 & inFFAb & No variability detected \\
    J223933-451414 & -87.3 & 64.8 & 99.1 & 78.0 & 125.3 & 132.5 & (in)FFAb & Variable spectral shape at $\leq$231\,MHz  \\
    J224408-202719 & -9.5 & 10.1 & 92.1 & 110.8 & 100.9 & 119.4 & inFFAb & Brightness change at $\leq$231\,MHz  \\
\hline
\end{tabular}%
\caption{The log likelihoods for each spectral model for each source. For each source, the best model was determined using the average Bayes factor to determine the most likely model over all epochs. See Section~\ref{sec:spectral_modelling} for details. Models are defined as: SSA, SSA with an exponential break (SSAb), external homogeneous FFA (FFA), external homogeneous FFA with a spectral break (FFAb), external inhomogeneous FFA (inFFA) and an external inhomogeneous FFA with a spectral break (inFFAb). A preferred model with an asterisk (*) next to it, indicates there was not strong evidence for this model compared to any other model (i.e. the Bayes factor was less than 100 for each pair of models). Preferred models with a (b) indicates that there is not strong evidence to support the presence of a high frequency spectral break over the absence of one but that the spectral model itself is preferred. Likewise, preferred models of (in)FFA indicate that an FFA model is preferred but there is not strong evidence of the FFA model over the inFFA model. Furthermore, a preferred model of (SSA)b indicates all spectral models with a high frequency spectral break are preferred but there is not strong evidence for the SSA model over any other spectral model. } 
\label{tab:targets_specmodels}
\end{table*}

\begin{figure*}
     \centering
     \begin{subfigure}[b]{0.49\linewidth}
         \centering
         \includegraphics[width=\linewidth]{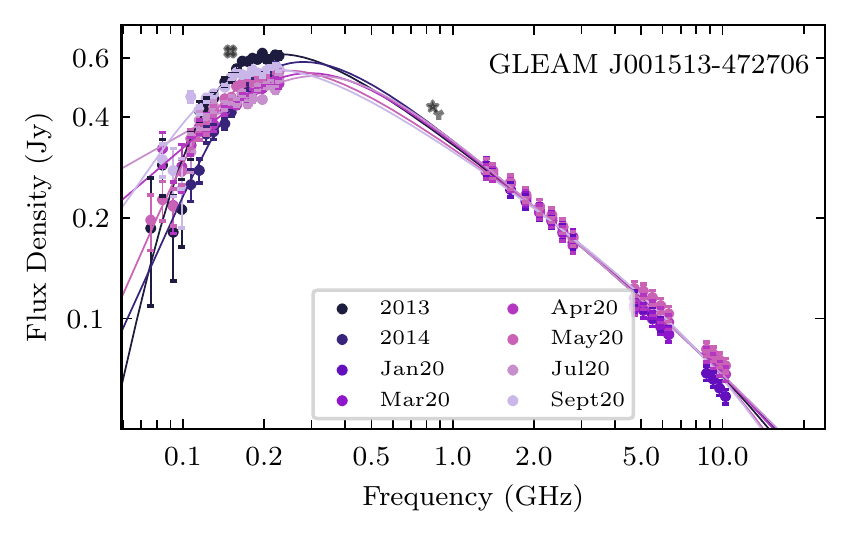}
         \vskip -0.5cm
         \caption{}
         \label{fig:001513_sed}
     \end{subfigure}
     \begin{subfigure}[b]{0.49\linewidth}
         \centering
         \includegraphics[width=\linewidth]{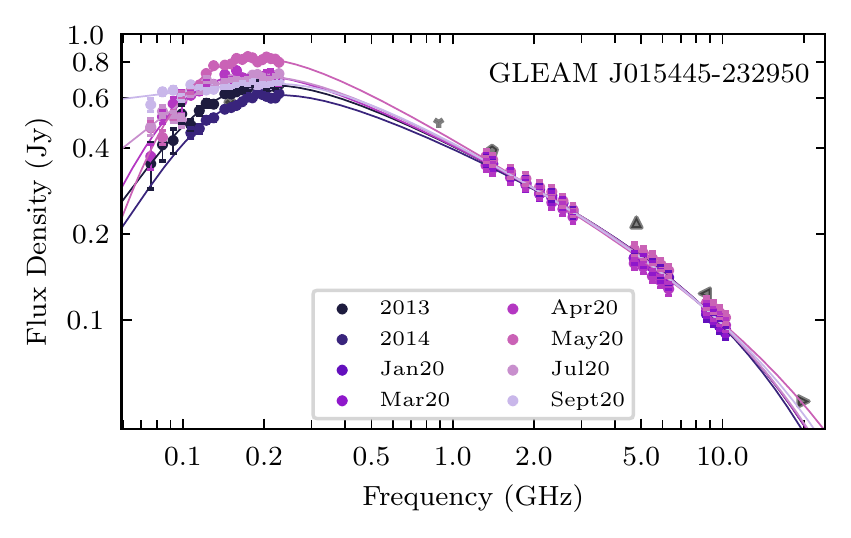}
         \vskip -0.5cm
         \caption{}
         \label{fig:015445_sed}
     \end{subfigure}
     \\
     \begin{subfigure}[b]{0.49\linewidth}
         \centering
         \includegraphics[width=\linewidth]{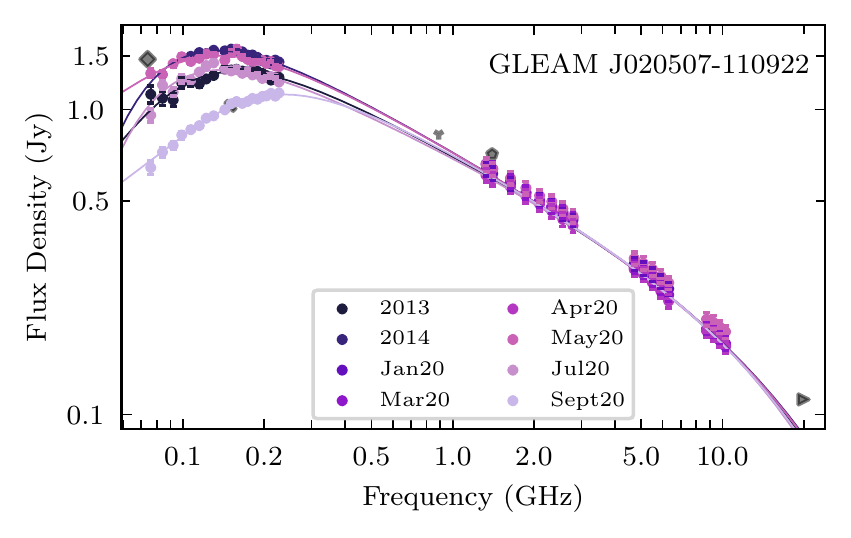}
         \vskip -0.5cm
         \caption{}
         \label{fig:020507_sed}
     \end{subfigure}
     \begin{subfigure}[b]{0.49\linewidth}
         \centering
         \includegraphics[width=\linewidth]{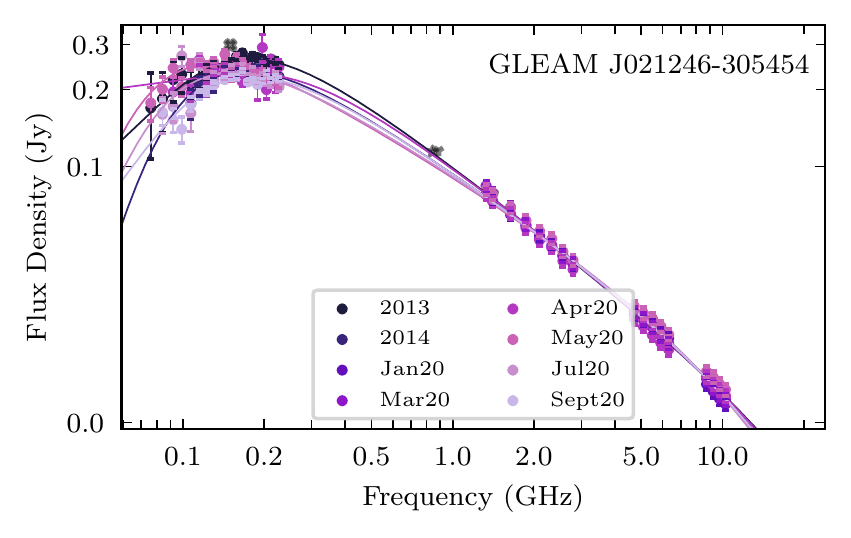}
         \vskip -0.5cm
         \caption{}
         \label{fig:021246_sed}
     \end{subfigure}
     \\
     \begin{subfigure}[b]{0.49\linewidth}
         \centering
         \includegraphics[width=\linewidth]{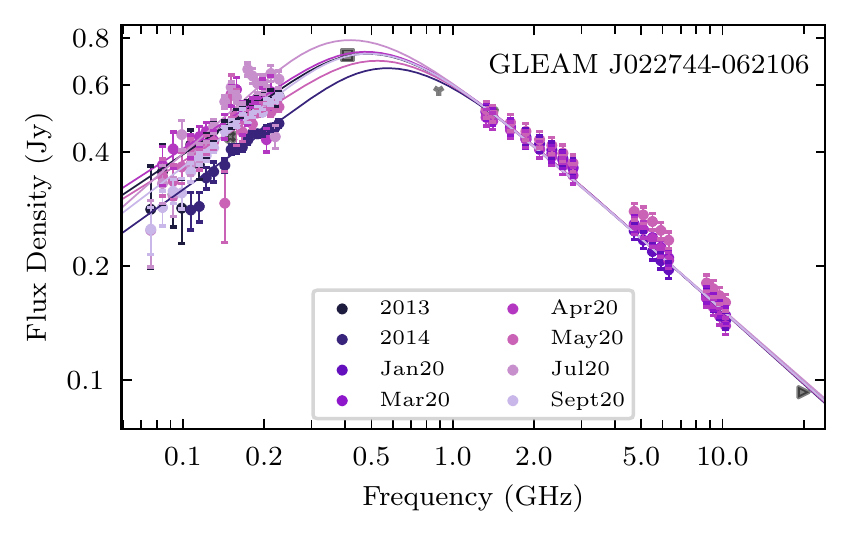}
         \vskip -0.5cm
         \caption{}
         \label{fig:022744_sed}
     \end{subfigure}
     \begin{subfigure}[b]{0.49\linewidth}
         \centering
         \includegraphics[width=\linewidth]{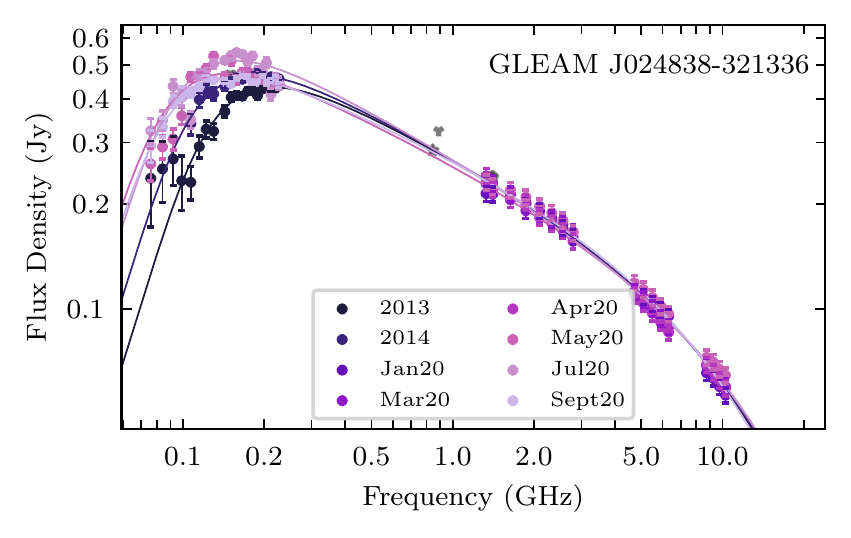}
         \vskip -0.5cm
         \caption{}
         \label{fig:024838_sed}
     \end{subfigure}
     \\
     \begin{subfigure}[b]{0.49\linewidth}
         \centering
         \includegraphics[width=\linewidth]{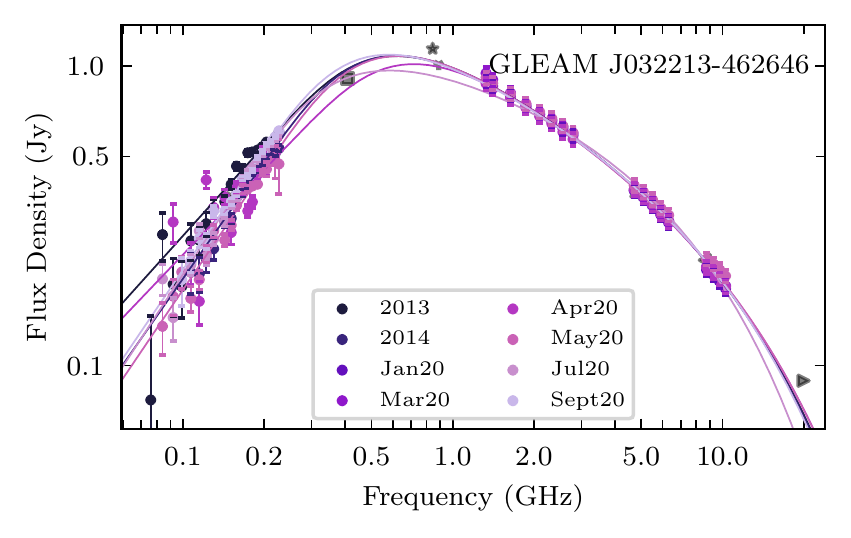}
         \vskip -0.5cm
         \caption{}
         \label{fig:032213_sed}
     \end{subfigure}
     \begin{subfigure}[b]{0.49\linewidth}
         \centering
         \includegraphics[width=\linewidth]{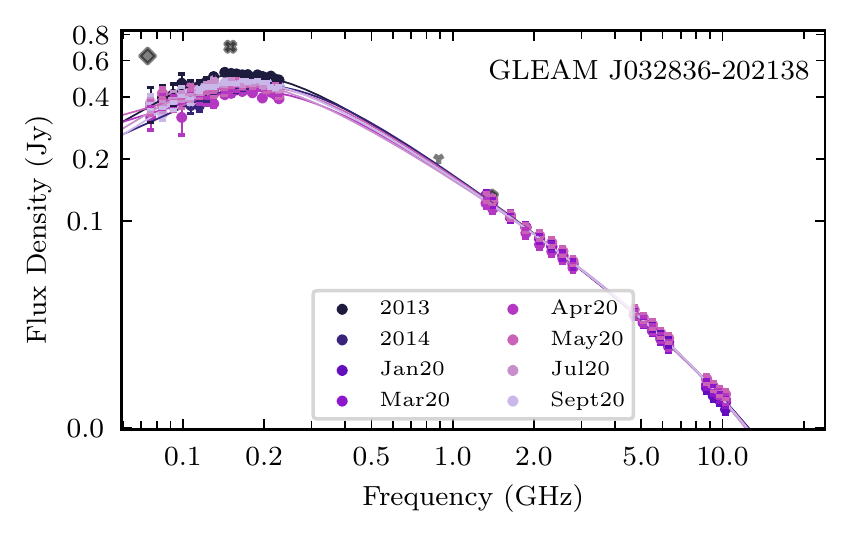}
         \vskip -0.5cm
         \caption{}
         \label{fig:032836_sed}
     \end{subfigure}
\caption{SEDs for all targets. ATCA and MWA data are plotted for each epoch with the best spectral model, according to the average Bayes Factor presented in Table~\ref{tab:targets_specmodels}, overlaid. Additional surveys are plotted in grey: VLSSr (diamond), TGSS-ADR1 (cross), MRC (square), SUMSS (star), RACS (Y), NVSS (pentagon), AT20G (20\,GHz: left arrow, 8.6\,GHz: right arrow, 4.8\,GHz upwards arrow)}
\label{fig:seds_pg1}
\end{figure*}

\begin{figure*}\ContinuedFloat
     \centering
     \begin{subfigure}[b]{0.49\linewidth}
         \centering
         \includegraphics[width=\linewidth]{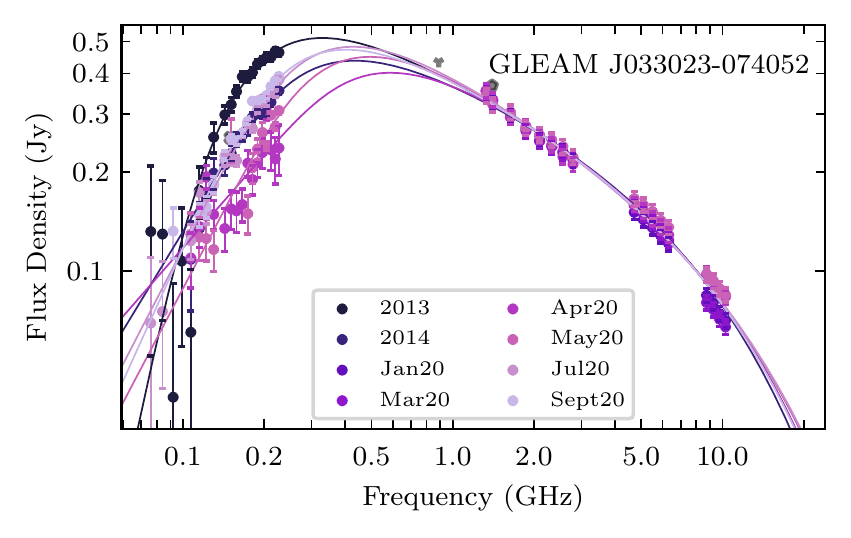}
         \vskip -0.5cm
         \caption{}
         \label{fig:033023_sed}
     \end{subfigure}
     \begin{subfigure}[b]{0.49\linewidth}
         \centering
         \includegraphics[width=\linewidth]{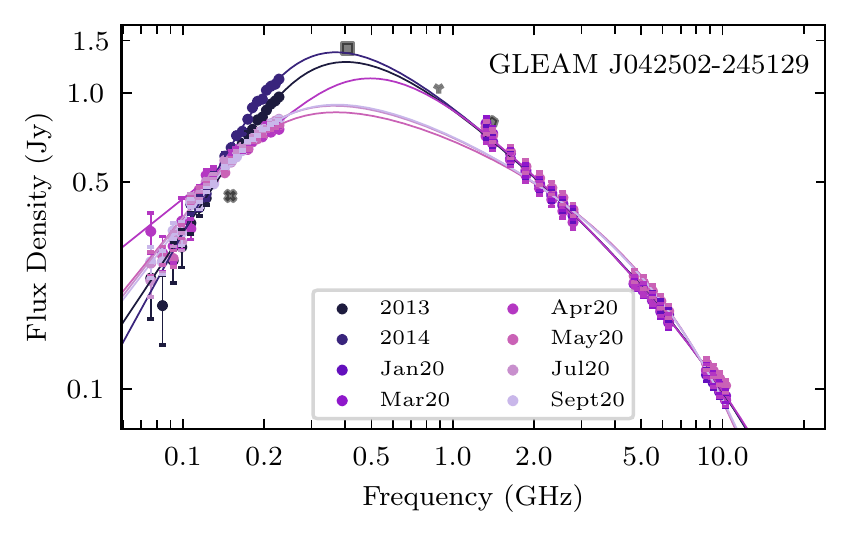}
         \vskip -0.5cm
         \caption{}
         \label{fig:042502_sed}
     \end{subfigure}
     \\
     \begin{subfigure}[b]{0.49\linewidth}
         \centering
         \includegraphics[width=\linewidth]{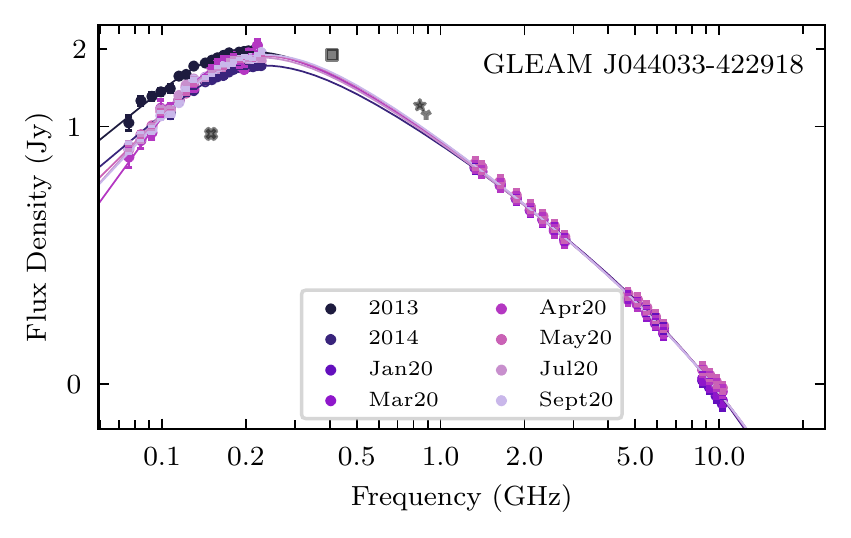}
         \vskip -0.5cm
         \caption{}
         \label{fig:044033_sed}
     \end{subfigure}
     \begin{subfigure}[b]{0.49\linewidth}
         \centering
         \includegraphics[width=\linewidth]{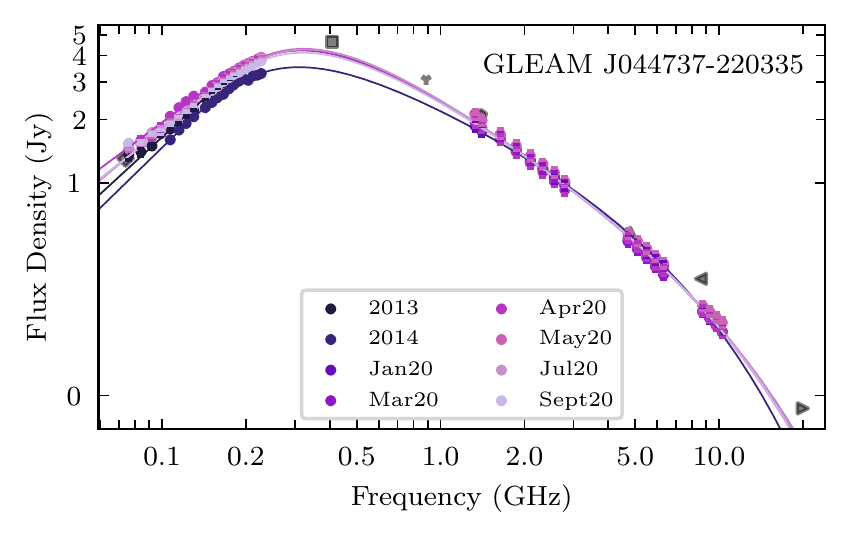}
         \vskip -0.5cm
         \caption{}
         \label{fig:044737_sed}
     \end{subfigure}
     \\
     \begin{subfigure}[b]{0.49\linewidth}
         \centering
         \includegraphics[width=\linewidth]{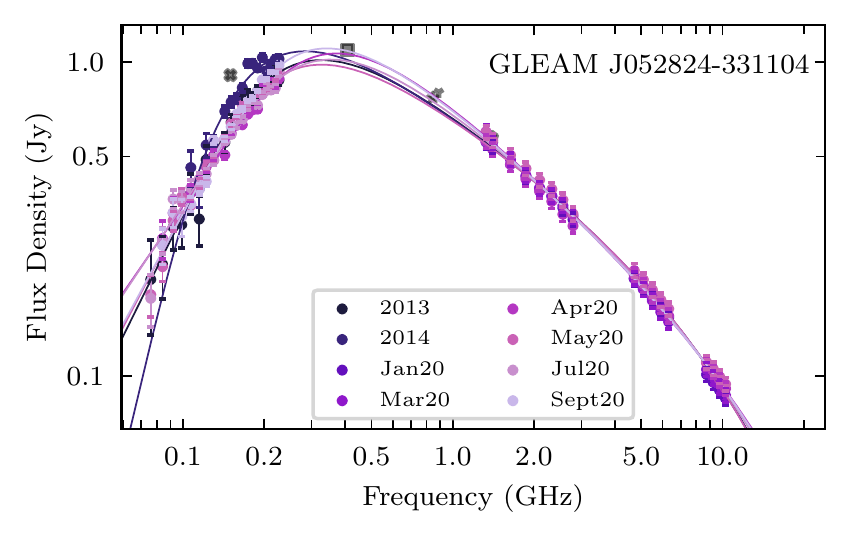}
         \vskip -0.5cm
         \caption{}
         \label{fig:052824_sed}
     \end{subfigure}
     \begin{subfigure}[b]{0.49\linewidth}
         \centering
         \includegraphics[width=\linewidth]{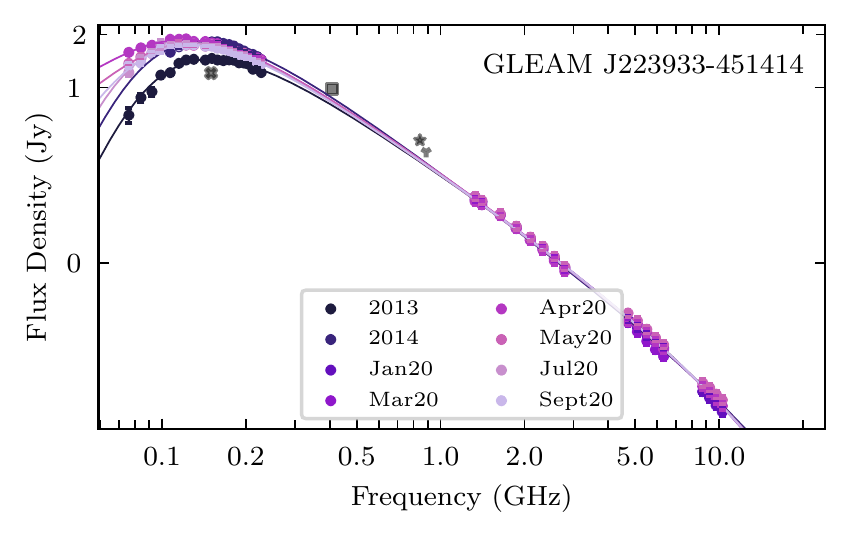}
         \vskip -0.5cm
         \caption{}
         \label{fig:223933_sed}
     \end{subfigure}
     \\
     \begin{subfigure}[b]{0.49\linewidth}
         \centering
         \includegraphics[width=\linewidth]{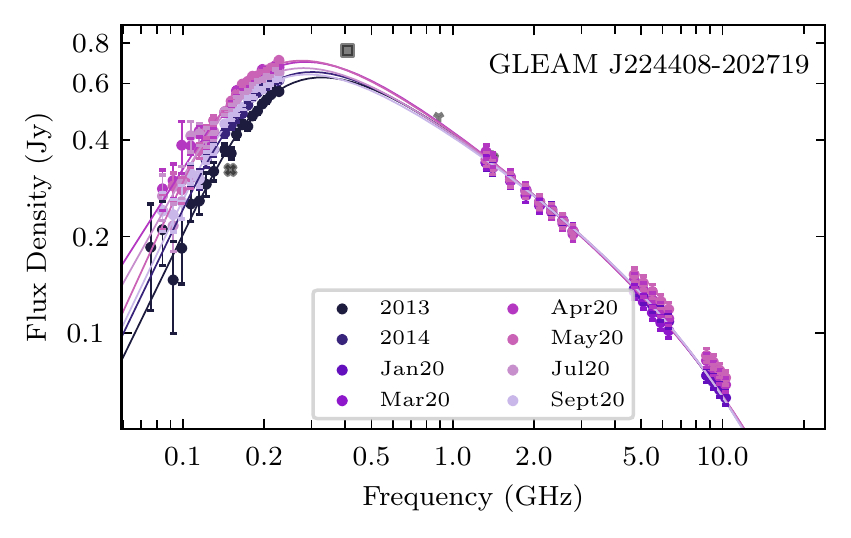}
         \vskip -0.5cm
         \caption{}
         \label{fig:224408_sed}
     \end{subfigure}
        \caption{(continued) SEDs for all targets. Models plotted are the best spectral model according to the average Bayes Factor presented in Table~\ref{tab:targets_specmodels}. Additional surveys are plotted in grey: VLSSr (diamond), TGSS-ADR1 (cross), MRC (square), SUMSS (star), RACS (Y), NVSS (pentagon), AT20G (20\,GHz: left arrow, 8.6\,GHz: right arrow, 4.8\,GHz upwards arrow)}
        \label{fig:seds_pg2}
\end{figure*}

\section{Results}
\label{sec:results}
We present the SEDs for each epoch of each source in Figure~\ref{fig:seds_pg1}. We include flux density measurements from other radio surveys in the SEDs; these were not used in any fitting, but are included in the plots for completeness. The additional radio surveys are the Very Large Array Low-frequency Sky Survey Redux \citep[VLSSr; ][]{2014MNRAS.440..327L}, Tata Institute for Fundamental Research Giant Metrewave Radio Telescope 150\,MHz Sky Survey Alternative Data Release 1 \citep[TGSS-ADR1; ][]{intema2016TGSS}\footnote{\citep[We use the rescaled version of TGSS by][ to match the GLEAM flux density scale]{2017arXiv170306635H}}, the Molonglo Reference Catalogue \citep[MRC; ][]{1981MNRAS.194..693L,1991Obs...111...72L}, Sydney University Molonglo Sky Survey \citep[SUMSS;][]{2003MNRAS.342.1117M}, Rapid ASKAP Continuum Survey \citep[RACS; ][]{2021PASA...38...58H}, NRAO VLA Sky Survey \citep[NVSS;][]{condon+98}, an the Australia Telescope 20\,GHz Survey \citep[AT20G; ][]{at20g}. All catalogues were cross-matched using \textsc{Topcat}'s \citep{2005ASPC..347...29T} nearest neighbour routine with a 2\,arcmin radius. A 2\,arcmin radius was chosen as it is comparable to the resolution of GLEAM. Table~\ref{tab:targets_specmodels} presents the results of the spectral fitting, reporting the average Bayes factor to determine the most likely spectral model over all epochs, and observed variability. 

We find each source shows a negative-slope power-law SED at frequencies $\geq1$\,GHz, which steepens at high frequency, consistent with synchrotron emission from a radio-loud AGN.
We do not find any sources in our sample with a flat spectrum at gigahertz frequencies. Furthermore, we do not detect any significant variability with the ATCA in the 2020 monitoring, which sampled timescales of up to four months. The ATCA spectra of targets were compared to their secondary calibrators and the bandpass calibrator, PKS\,B1934--638, but no target showed significant variability or trends. 

In contrast, there are several different behaviours of variability detected at megahertz frequencies. Most common are sources showing a consistent spectral shape with small variations with an overall trend over the epochs. However, some appear to change their spectral shape significantly, for example GLEAM\,J020507--110922 shown in Figure~\ref{fig:020507_sed}. For each source, we check for significant variability with the MWA by comparing to nearby ($\lesssim1$\,deg) sources. In each case, we find no significant variability or common behaviours between our targets and nearby sources. The SEDs of nearby targets can be found in the online supplementary materials. 

There is no significant variability at any frequency or timescale for GLEAM\,J052824--331104 or GLEAM\,J022744--062106. We do not detect any greater difference in flux densities for either source compared to any nearby source in the MWA images. GLEAM\,J052824--331104 was in the region of the SGP mosaics that R21 deemed too poor quality to detect variability (see Section~\ref{sec:obs_srces} for details). We therefore conclude any difference observed between 2013 and 2014 for GLEAM\,J052824--331104 is not physical. It is possible the initial variability of GLEAM\,J022744--062106 detected by R21 was genuine and not due to introduced instrumentation errors, and it was still variable in 2020. However, the noise of the images created in 2020 were of too low quality to detect any significant changes in flux density as the Sun was in the primary beam for several images.

The SSA model with a spectral break was the most likely model for only one source, GLEAM\,J024838--321336. However, the Bayes factors for the SSAb model compared to the next most likely models, the FFAb and inFFAb, are 33 and 90, respectively. Consequently, there is not strong evidence to support the SSAb spectral model over either the FFAb or inFFAb, given a $K>100$ is considered strong evidence. A higher frequency spectral break is more likely in each case, but there is low evidence for distinguishing between the spectral models: SSA, FFA or inFFA. This is likely due to the shifting peak frequency (shown in Figure~\ref{fig:024838_nu_p} and discussed further in Section~\ref{sec:discussion_j024838}) and insufficient sampling below the spectral peak (since $\nu_p \leq 140$\,MHz in all epochs). 

The other 14 sources were best fit with an inhomogeneous FFA spectral model with a spectral break. For GLEAM\,J001513--472706, the Bayes factor for inFFAb compared to the second most likely model, inFFA, is 82. This also suggests there is not enough strong evidence to support the inFFAb model, however, the Bayes factor for either inhomogeneous FFA model compared to either the FFA or SSA models is $\gg$100. We therefore conclude that the spectrum of GLEAM\,J001513--472706 is best fit by an inhomogeneous FFA model but the presence of a exponential break is uncertain. This is likely due to the lack of higher frequency flux densities at 9\,GHz in March~2020, which is roughly the frequency where we could expect a spectral break.

Lastly, we note GLEAM\,J021246--305454 has a Bayes factor of 5.7 when comparing the inFFAb model with the SSAb model (the second most likely spectral model). This is not decisive evidence, so we cannot confidently say the inFFAb spectral model is the most appropriate. Comparing the log likelihoods for each model presented in Table~\ref{tab:targets_specmodels}, there is strong evidence GLEAM\,J021246--305454 has an exponential spectral break. However, similar to GLEAM\,J024838--321336, there is low evidence to distinguish between the spectral models. Again, it is likely this is due to insufficient sampling below the spectral turnover ($\approx$150\,MHz).

\section{Discussion}
\label{sec:discussion}
In this section, we will discuss the likely physical mechanisms for any observed variability. The majority of sources appear to show slow trends of increasing or decreasing flux density across the MWA band throughout 2020 with no significant variability detected with the ATCA. 

In Section~\ref{sec:discussion_scintillators}, we present the sources that are likely showing variability due to interstellar scintillation and discuss the implications of such a mechanism. We focus on individual sources that show uncommon variability; \ffasource{}, GLEAM\,J024838--321336, GLEAM\,J015445--232950 and GLEAM\,J223933--451414 in Sections~\ref{sec:discussion_j020507}, \ref{sec:discussion_j024838}, \ref{sec:discussion_j015445} and \ref{sec:discussion_j223933}, respectively.

\subsection{Interstellar Scintillation}
\label{sec:discussion_scintillators}
The large spectral coverage of these observations samples the two different regimes of scattering: weak and strong. The electron column density along the line of sight and observing frequency determine which scattering regime is applicable \citep{narayan1992physics,1998MNRAS.294..307W}. The electron column density is largely related to the Galactic latitude. All our sources are far away from the Galactic plane, thus the transition frequency, $\nu_0$, from strong to weak scattering is $\sim$8\,GHz and the angular size limit at the transition frequency, $\theta_{F0}$, is 4\,$\mu$as \citep{1998MNRAS.294..307W}. Continuing under this assumption, all our calculations for ISS at 2.1\,GHz, 5.5\,GHz and megahertz frequencies will be using the strong scattering regime, while the 9\,GHz calculations will be using the weak scattering regime. 
Furthermore, we eliminate the possibility of diffractive ISS in the strong regime, as the fractional bandwidth of variations is predicted to be $\sim1.4\times10^{-5}$, but the smooth SED for all sources at frequencies $<8$\,GHz, in each epoch suggests the fractional bandwidth is closer to unity. 

\subsubsection{Weak Scattering}
First, we consider the modulation and timescales of variability due to weak scattering at 9\,GHz for a compact source. A compact source is defined as having angular size $\leq\theta_F$ where: 
\begin{equation}
    \theta_F = \theta_{F0}\sqrt{\frac{\nu_0}{\nu}},
    \label{eq:weak_thetaf}
\end{equation}
\noindent
resulting in a timescale of scintillation according to: 
\begin{equation}
    t_{\mathrm{compact}} \approx 2 \sqrt{\frac{\nu_0}{\nu}},
    \label{eq:weak_tf}
\end{equation}
where $\theta_{F0}$ is the angular size limit of a source at an observation frequency, $\nu$ that equals the transition frequency, $\nu_0\approx8$\,GHz at our Galactic latitude \citep{1998MNRAS.294..307W}.

Using Equation~\ref{eq:weak_tf}, for observations observed at frequency, $\nu$, of 9\,GHz, the timescale of modulation due to ISS, $t_\mathrm{compact}$, would be of the order of 1.9\,hours. Any observations over several hours would therefore average over the variability due to ISS. All our observations for ATCA were taken over observations blocks of $\sim18$\,hours, thus our measured flux densities average over any hourly variability. Thus, no significant variability would be detected in our observations. 

\begin{figure}
    \centering
    \includegraphics[width=\linewidth]{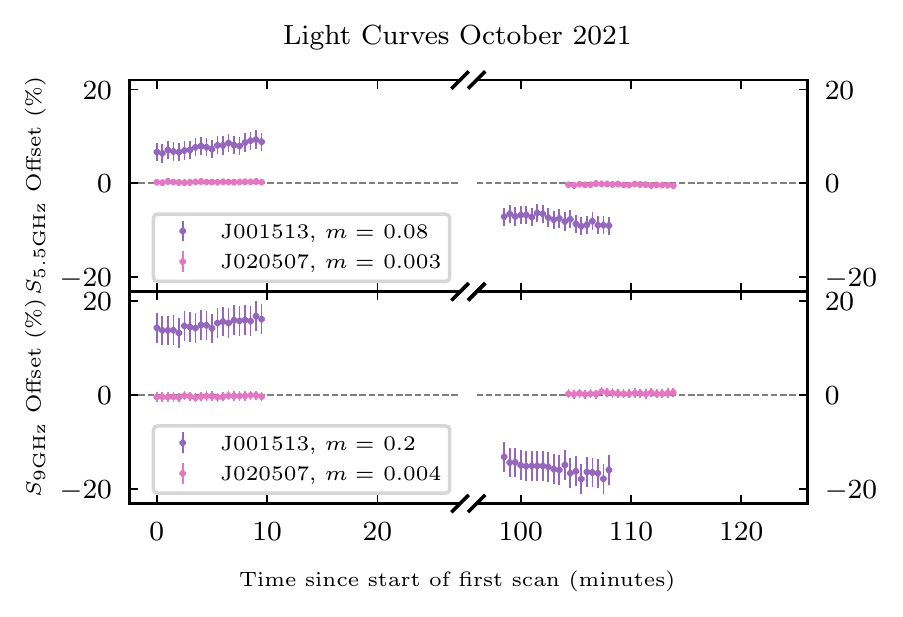}
    \caption{Light curves of flux density variance for GLEAM\,J001513--472705 (purple) and GLEAM\,J020507--110922 (pink) in October~2021 at 5\,GHz (top) and 9\,GHz (bottom). Flux densities were measured using the \texttt{uvmodelfit} function of \textsc{casa} for 30\,second time intervals. The fractional flux density percentage offset (or modulation, $m$) was calculated as the difference of each 30\,second flux density measurement from the median flux density of the entire light curve for each source. The errors on the flux density are the local rms of the images for each source only, as any systematic errors do not influence the fractional flux density offset. }
    \label{fig:lightcurves}
\end{figure}

To test this hypothesis, we analyse the ATCA Director's Time data collected hourly in October~2021, see Section~\ref{sec:obs_atca}. The October~2021 follow-up observations with the ATCA were taken using a different observing technique to the original 2020 monitoring. These observations consisted of multiple 10\,minute scans separated by a couple of hours. Let us take GLEAM\,J001513--472705 as an example source for future calculations. In the October~2021 epoch, we observed GLEAM\,J001513--472705 twice with 10\,minute scans separated by $\sim 1.5\,$hours, which is slightly below the expected timescale of 1.9\,hours. Figure~\ref{fig:lightcurves} presents the light-curves of GLEAM\,J001513--472705 in October~2021 at 5 and 9\,GHz. Flux density measurements were taken at 30\,second intervals in $(u,v)$ space using the \texttt{uvmodelfit} module in \textsc{casa} and the percentage offset is calculated from a median flux density value\footnote{In each case, our target dominates the visibilities ensuring such model fitting is appropriate.}. 

Within the 10\,minute scans, there may be modulation (seen as rising in the first scan and then decreasing in the second) but it is likely this is sampling a small fraction of the longer-term (hourly) modulation. In the 2021 observations we see an overall modulation of $\approx 0.15$ at 9\,GHz.
We can calculate the expected modulation using: 
\begin{equation}
    m_\mathrm{compact} = \left({\frac{\nu_0}{\nu}}\right)^{17/12},
    \label{eq:weak_mp}
\end{equation}
\noindent
which suggests $m_\mathrm{compact}\approx0.85$. It is worth noting, Equation~\ref{eq:weak_mp} applies for a well sampled light-curve, since we only have poor time sampling, this calculated modulation has a large margin of error. The smaller measured modulation could be due to a number of factors: we are not sampling the entire timescale or modulation of variability, and/or the source is slightly resolved compared to the angular size limit, $\theta_{F0}$. The October~2021 observations only consisted of two 10\,minute scans, hence it is likely that the modulation and timescale is not sampled sufficiently. Further observations of GLEAM\,J001513--472705 at 9\,GHz with continued monitoring over timescales of hours to days would increase the likelihood of sampling the entire timescale of variability and converging on the modulation.  

Alternatively, if GLEAM\,J001513--472705 has a compact component that is slightly resolved compared to the angular size of the scattering screen, the modulation decreases and timescale increases according to: 
\begin{equation}
    m_{\mathrm{observed}} = m_\mathrm{compact} \left({\frac{\theta_F}{\theta_S}}\right)^{7/6},
    \label{eq:weak_mobs}
\end{equation}
\begin{equation}
    t_{\mathrm{observed}} = t_\mathrm{compact} \left({\frac{\theta_S}{\theta_F}}\right),
    \label{eq:weak_tobs}
\end{equation}
\noindent
where $\theta_F$ is defined by Equation~\ref{eq:weak_thetaf} and is the angular scale of the scattering screen and $\theta_S$ is the angular size of the compact component \citep{1998MNRAS.294..307W}. If we have sampled the timescale and modulation sufficiently, GLEAM\,J001513--472705 must have a compact component $\approx 17$\,$\mu$as. This would correspond to a timescale of scintillation of roughly 8.5\,hours. It is likely we over overestimated the modulation and underestimated the timescale based on our poor sampling, as such, this compact component size estimate should be considered as a lower limit.

These caveats to the weak scattering are reasonable assumptions to explain the variability of GLEAM\,J001513--472705 measured at 9\,GHz. This would imply that GLEAM\,J001513--472705 is an intra-day variable source with a compact feature on $\mu$as scales. Further monitoring at 9\,GHz would be required to sample the modulation more thoroughly and estimate the timescales of ISS more accurately. 

\subsubsection{Strong Scattering}
Let us now consider whether any variability at frequencies $<\nu_0$ are also consistent with interstellar scintillation but in the strong regime, in particular due to refractive interstellar scintillation (RISS). In the strong regime, we have:
\begin{equation}
    \theta_r = \theta_{F0}\left(\frac{\nu_0}{\nu}\right)^{11/5},
    \label{eq:strong_thetar}
\end{equation}
\begin{equation}
    m_\mathrm{compact} = \left({\frac{\nu}{\nu_0}}\right)^{17/30}, m_{\mathrm{observed}} = m_\mathrm{compact} \left({\frac{\theta_r}{\theta_S}}\right)^{7/6}
    \label{eq:strong_mp}
\end{equation}
\begin{equation}
    t_\mathrm{compact} = 2\left({\frac{\nu}{\nu_0}}\right)^{11/5}, t_{\mathrm{observed}} = t_\mathrm{compact} \left({\frac{\theta_S}{\theta_r}}\right)
    \label{eq:strong_tf}
\end{equation}
\noindent 
following \citet{1998MNRAS.294..307W}, where a compact source is defined as $\leq\theta_r$. 

At 5\,GHz, we would expect a modulation of $\sim$0.77 on timescales of $\sim$6\,hours with a angular screen size, i.e. the angular size of a compact component, of $\sim$11.2\,$\mu$as. Considering GLEAM\,J001513--472705 as an example again, we measure a modulation at 5\,GHz of $\sim$0.086 across approximately two hours. Consistent with the results of the 9\,GHz variability, this calculation suggests that the compact feature of GLEAM\,J001513--472705 is likely resolved compared to the scattering screen and/or we have not sampled the timescale and modulation sufficiently.

Furthermore, at 150\,MHz, using Equation~\ref{eq:strong_tf}, the timescale of variability is expected to be 1.4\,years with a modulation of 0.1 (using Equation~\ref{eq:strong_mp}) and a scattering screen angular size $\approx 25$\,mas (using Equation~\ref{eq:strong_thetar}).  Our observations during 2020 cover a timescale of six months with four epochs. Thus, we should be able to detect a small level of variability as a slow shift in flux density across the entire MWA band over the course of the observations. For GLEAM\,J001513--472705, we see a modulation of 0.1 over the 6\,month monitoring period with a constant trend of the flux density increasing across the entire MWA band. Several other sources also display slow trends of increasing/decreasing flux density across the entire MWA band in the 2020 observations: GLEAM\,J021246--305454, GLEAM\,J032213--462646, GLEAM\,J032836--202138, GLEAM\,J033023--074052, GLEAM\,J042502--245129, GLEAM\,J044033--422918, GLEAM\,J044737--220335, GLEAM\,J224408--202719. Since the variability detected for GLEAM\,J001513--472705 at each frequency band is consistent with ISS, it is likely the sources that show a similar variability trend at MHz frequencies are also variable due to ISS. GLEAM\,J015445--232950 and \ffasource{} also show trends of variations in the flux density across the MWA band. However, both also display a change in their spectral shape within the MWA band in later epochs. We discuss the variability of \ffasource{} and GLEAM\,J015445--232950 further in Section~\ref{sec:discussion_j020507} and Section~\ref{sec:discussion_j015445} respectively.

It is worth noting, R21 suggest sources with a low MOSS value ($<36.7$) are likely variable due to refractive ISS. In agreement with R21, of our 15 targets, we find all sources with a low MOSS value to be exhibiting variability consistent with ISS apart from one source which shows no significant variability (GLEAM\,J022744--062106). Inversely, inspecting the MOSS value of the 9 sources we claim are exhibiting ISS, all bar one (GLEAM\,J033023--074052) have a low MOSS value consistent with ISS according to R21. 

We would thus expect these sources to show intra-day variability at higher frequencies $>1$\,GHz. While it is uncommon for PS sources to have hot-spots or compact features in their morphologies \citep{keim2019extragalactic}, 9 of our 15 PS sources show variability entirely consistent with scintillation due to such a compact feature. High-resolution imaging would determine the presence of a compact feature on $\mu$as to mas scales.

\subsection{GLEAM J020507--110922}
\label{sec:discussion_j020507}
\begin{figure}
    \centering
    \includegraphics[width=\linewidth]{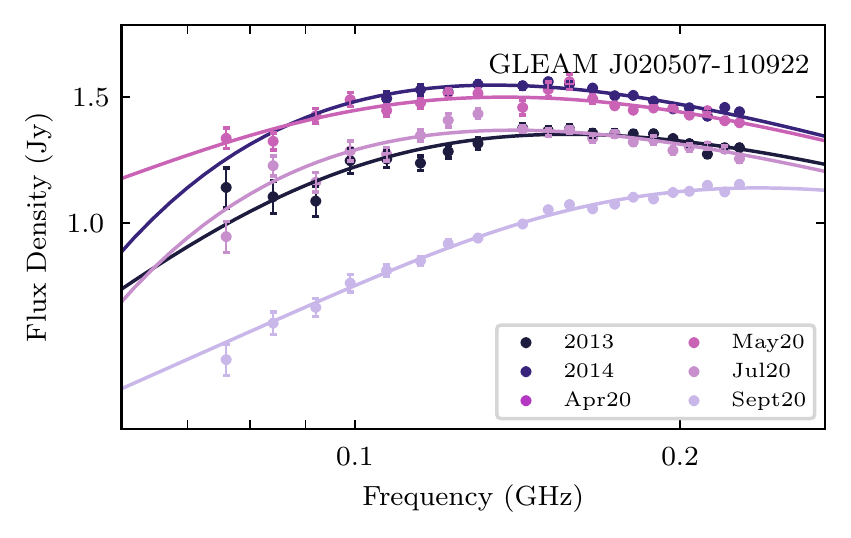}
    \caption{SED of only MWA flux densities for GLEAM\,J020507--110922. The spectral shape in September~2020 (lavender) is significantly different to all previous epochs. }
    \label{fig:020507_mwa}
\end{figure}

Due to the unique and extreme nature of the variability exhibited by \ffasource{}, we discuss several plausible explanations: intrinsic variability due to SSA, ISS, and variations in the free-free opacity. A close-up of variability observed at megahertz-frequencies for \ffasource{} is presented in Figure~\ref{fig:020507_mwa}.

\subsubsection{Synchrotron Self Absorption}
\begin{figure}
    \centering
    \includegraphics[width=0.9\linewidth]{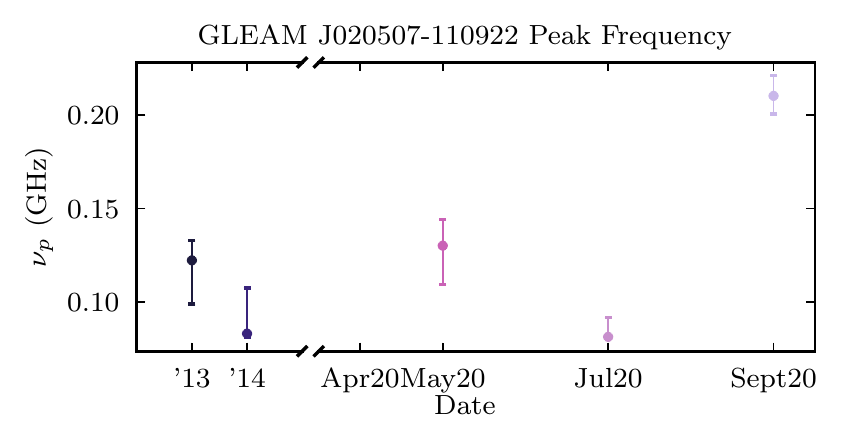}
    \caption{The measured peak frequency ($\nu_\mathrm{p}$) from the fitted spectral model (inFFAb) of GLEAM\,J020507--110922 for each epoch. $\nu_\mathrm{p}$ is consistent from 2013 to July~2020 and then significantly increases in September~2020.}
    \label{fig:j020507_freqpeakvstime}
\end{figure}

Firstly, we assume that the mechanism for the turnover in \ffasource{} is due to SSA. Any changes in flux density below or around the turnover would be due to changes in the synchrotron absorption. Using a synchrotron model, with $m_e$ and $e$ the electron mass and electron charge respectively, we have (in the observed frame of reference),
\begin{equation}
    S_{\nu_p} = \left( \frac{\pi^3 m_e^3\nu_p^5 \theta_S^4}{0.94 e B \sin(\theta)} \right)^{\frac{1}{2}},
    \label{eq:ssa_peakflux}
\end{equation}
\noindent
where $\theta_S$ is the angular source size, and the magnetic field, $B$, is at an angle $\theta$ to the line of sight \citep{tingay2015spectral}. Thus, changes in the peak frequency, $\nu_p$, would result in changes to the flux density at the peak frequency, $S_{\nu_p}, \propto \nu_p^{5/2}$. Using the best model fit for each epoch, Figure~\ref{fig:j020507_freqpeakvstime} shows the change in $\nu_p$ with time. Therefore, assuming a constant $\theta_S$, $B$ and $\sin(\theta)$, the measured change in $\nu_p$ of ~0.1\,GHz would correspond to $S_{\nu_p}$ increasing by $\approx4$\,mJy. However, we detect a \textit{decrease} in $S_{\nu_p}$ of $\approx0.5$\,Jy. Either the magnetic field would need to increase by several orders of magnitude, or the source size would need to contract significantly ($\sim10$\%); both scenarios are physically improbable. Consequently, we can eliminate the possibility that the variability is due to variations of the synchrotron emission. 

\subsubsection{Interstellar Scintillation}
Secondly, we consider the possibility of ISS following the same calculations described in Section~\ref{sec:discussion_scintillators}. As the 9\,GHz data is the only frequency in the weak regime, we start by examining these data. We do not detect any modulation at 9\,GHz on timescales of months or years, however, as described in Section~\ref{sec:discussion_scintillators}, we would expect weak scattering on the timescale of $\sim $1.9\,hours. The October~2021 observations of \ffasource{} consisted of two 10\,minute scans separated by $\sim$2\,hours. Thus, we can expect any modulations we detect between the scans to be due to weak scattering. At 9\,GHz, we measure a modulation of $\approx 0.0063$, which is two orders of magnitude smaller than the expected modulation of 0.85 at 9\,GHz. There is no variability detected by eye in the light curve for \ffasource{} presented in Figure~\ref{fig:lightcurves}. Consequently, \ffasource{} would have to be resolved compared to the scattering screen. If the variability observed at 9\,GHz between the 10\,minute scans is due to weak scattering, \ffasource{} must have a compact component of the scale 0.3\,mas. Furthermore, it would increase the timescale of observed variability to five days. While it is unlikely \ffasource{} has a compact feature $<1$\,mas, we recommend monitoring over the course of several consecutive days at 9\,GHz to confirm. 

Next, we consider the variability in the strong regime at 5\,GHz. We detect no significant variability by eye at 5\,GHz in the October~2021 observations and calculate a modulation of only 0.0025 at 5\,GHz between the 10\,minute scans in October~2021. Again, this significantly smaller modulation at 5\,GHz could be explained by \ffasource{} being slightly resolved compared to the refracting scintillating screen, but there is still a constraint on the compact component of $\leq$1.5\,mas. The typical power-law spectrum of \ffasource{} above the turnover suggests it is unlikely there is a compact component contributing a large fraction of the flux density at 5\,GHz and 9\,GHz that is smaller than 2\,mas. Furthermore, this modulation is well within the $1\sigma$ flux density errors of \ffasource{}: thus, no significant modulation is detected between the 10-minute scans. 

Continuing with considering the strong regime but now at 150\,MHz, we find there is a noticeably different spectral shape in September~2020 compared to previous epochs, see Figure~\ref{fig:020507_mwa}. Such a change in spectral shape would require small scale structures within the refracting screen creating a frequency dependence smaller than the bandwidth of the MWA. While it is not impossible, it is unlikely that such small scale structures only appeared between July and September~2020. The constant spectral shape until the September~2020 epoch suggests a different physical mechanism may have caused the observed variability between July and September~2020. 
We will therefore consider the variability of the other epochs and exclude September~2020 first. 

At 150\,MHz, the timescale of variability is expected to be $\sim1.4$\,years according to Equation~\ref{eq:strong_tf}. We see consistent variability between the epochs of observation on scales of months, suggesting that \ffasource{} must be entirely compact compared to the refracting plasma at 150\,MHz. As shown in Section~\ref{sec:discussion_scintillators}, the scale of the scattering disc is $\approx 25$\,mas at 150\,MHz. It is possible \ffasource{} has a compact component $\sim$25\,mas in size that is dominating the flux density measured at 150\,MHz; i.e. that the resolved lobes are contributing a small, almost negligible, portion of the flux density at MHz frequencies or that \ffasource{} is extremely compact. Therefore, the variability observed by the MWA is possibly due to RISS, provided \ffasource{} is $\sim$25\,mas. Furthermore, there would need to be small-scale structures ($<\theta_r = 25$\,mas) in the scintillating screen inducing strong frequency dependence between July and September~2020. Such small structures in the plasma would be comparable to the scales of plasma required for an extreme scattering event (ESE). ESEs are rare events and high-quality dynamic radio spectra are required to characterise the features of the plasma causing such an event \citep{Bannister2016ESEs,2016ApJ...817..176T}.

High resolution images using VLBI would be able to confirm or deny the presence of a scintillating compact feature. The high resolution images paired with continued monitoring at MHz frequencies (on timescales of $\sim$years) and GHz frequencies (on timescales of $\sim$days) would be able to determine the dominance of the compact feature and morphology at multiple frequencies. 

\subsubsection{Variable Optical Depth}
\begin{figure}
    \centering
    \includegraphics[width=\linewidth]{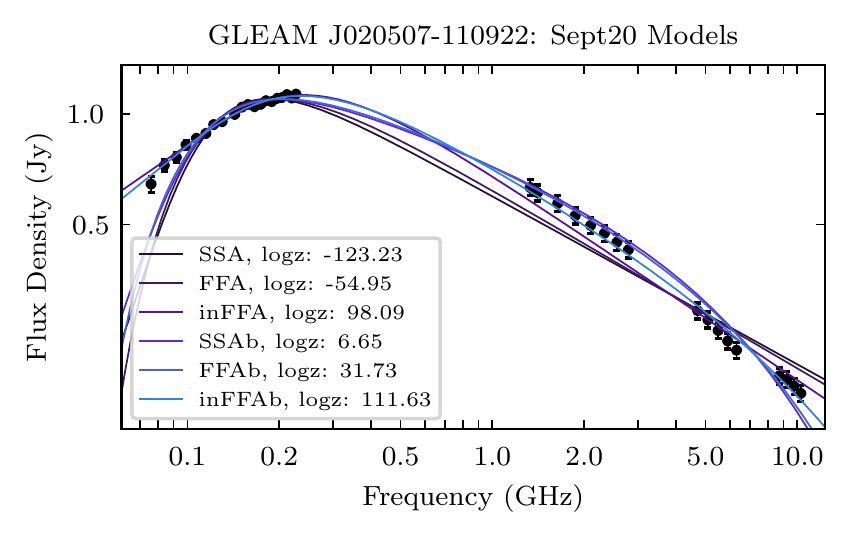}
    \caption{SED for GLEAM\,J020507--110922 using the MWA September~2020 epoch and the average ATCA flux densities used for spectral fitting. The models plotted are the six spectral models fitted to the September~2020 flux densities. The log likelihood of each spectral model is presented in the legend, a higher log likelihood suggests more evidence for the spectral model. All spectral models other than an inhomogenenous free-free absorption spectral model are inadequate at explaining the flux densities below the spectral turnover. }
    \label{fig:j020507_octmodels}
\end{figure}

Lastly, we consider the possibility that the variability is due to variations in the optical depth of an ionised plasma screen. If we assume all the variability seen at 100\,MHz is due to variations in this optical depth, we can scale the variations up to 5\,GHz and 9\,GHz as the free-free opacity, $\tau_\mathrm{ff}$, scales according to $\nu^{-2.1}$ \citep{lang2013astrophysical}. We see a flux density change at 100\,MHz of 0.7\,Jy, which would scale to variations of 0.2\,mJy at 5\,GHz and 0.05\,mJy at 9\,GHz. Both these are well within the measurement error on the flux density measurements of \ffasource{} at 5\,GHz and 9\,GHz, suggesting inhomogeneities in the free-free absorbing media are consistent with the variability seen at all frequencies. Continuing under this assumption, we can calculate the opacity change, $\Delta\tau_\mathrm{ff}$, according to: 
\begin{equation}
    \Delta \tau_{\mathrm{ff}} = - \ln\left[{1 - \frac{\Delta S}{S_0 e^{-\tau_\mathrm{ff}}}}\right],
    \label{eq:opacity_variation}
\end{equation}
where $\Delta S$ is the change in flux density, and $S_0$ is the flux density of the compact region \citep{tingay2015spectral}. We calculate an upper limit on the opacity variation (by setting $\tau_\mathrm{ff}$ to 0), using the median flux density at 100\,MHz of 1.2\,Jy as $S_0$, of $\Delta\tau_{\mathrm{ff}}<0.88$. This suggests a large density gradient within the free-free cloud. The optical depth due to FFA is proportional to the electron temperature and free electron density, thus changes in either would result in changes to the overall absorption \citep{1997ApJ...485..112B}. It is possible a region in the free-free absorbing cloud with a higher density of free electrons or a ``clump'' with a lower electron temperature moved into the line of sight between July~2020 and September~2020. As the optical depth is proportional to the emission measure, $EM$, and electron temperature, $T_e$ according to $EM\times T_e^{-1.35}$ \citep{1967ApJ...147..471M}, we can calculate the ratio of the optical depth in September~2020 to July~2020. We find the $EM\times T_e^{-1.35}$ in September~2020 is $\sim$7.42 times that of July~2020. This would explain the significant change in spectral shape from July to September~2020. It is also worth noting the September~2020 epoch is inconsistent with all spectral models except an inhomogeneous free-free absorbing model with an exponential break at higher frequencies, shown in Figure~\ref{fig:j020507_octmodels}. The consistency with an inhomogeneous free-free absorbing model is consistent with the explanation of a denser or cooler region in the inhomogeneous surrounding cloud changing the optical depth at megahertz frequencies. 

To summarise, the variability of \ffasource{} is inconsistent with changes in the synchrotron emission and DISS. While it is possible to explain the majority of variability with ISS, it requires extreme constraints on the source size of $<2$\,mas above 5\,GHz and $<25$\,mas at 150\,MHz and small scale structures within the scattering screen. Changes in the optical depth can explain all of the variability seen at MHz frequencies and the insignificant variability seen in the GHz regime, as well as the change in spectral shape between July and September~2020. Furthermore, the spectral SED of \ffasource{} in September~2020 is best described by an inhomogeneous free-free absorbing model, consistent with the variability being explained by inhomogeneities in the free-free absorbing media.

\subsection{GLEAM J024838--321336}
\label{sec:discussion_j024838}

\begin{figure}
    \centering
    \includegraphics[width=\linewidth]{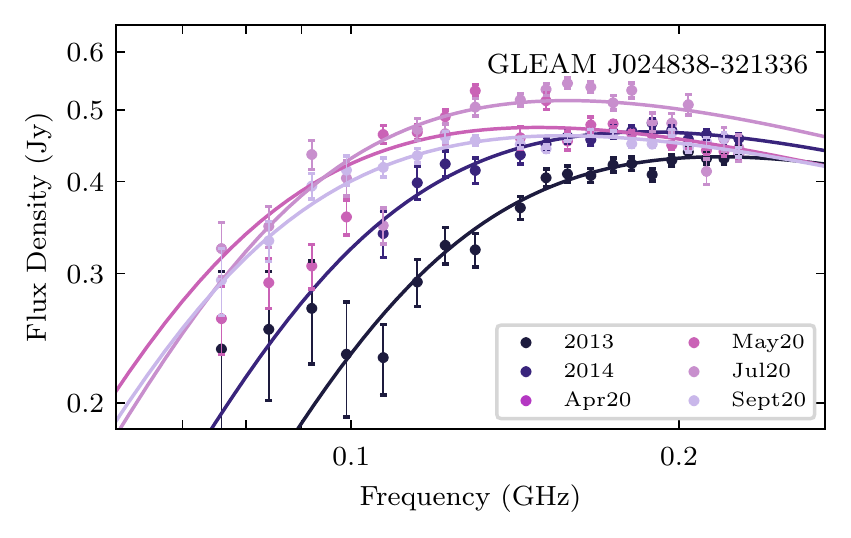}
    \caption{SED of only MWA flux densities for GLEAM\,J024838--321336. The similar spectral shape but shifting peak frequency to lower frequencies is consistent with an ejection cooling and adiabatically expanding as it travels across the jet. }
    \label{fig:024838_mwa}
\end{figure}

\begin{figure}
    \centering
    \includegraphics[width=\linewidth]{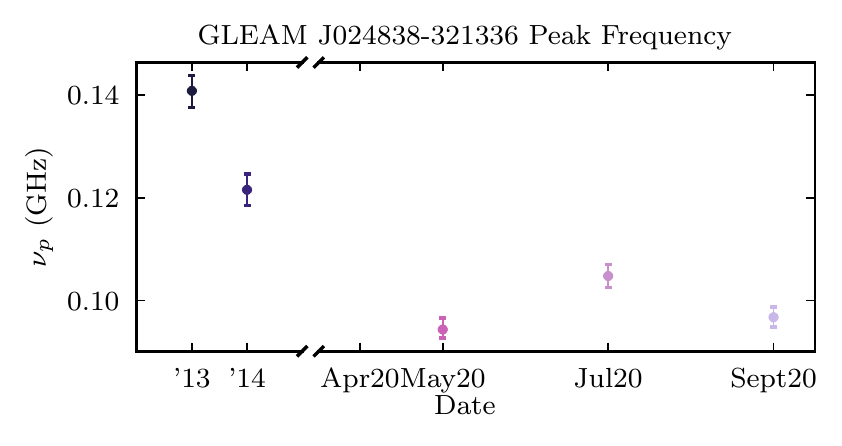}
    \caption{The measured peak frequency ($\nu_\mathrm{p}$) from the fitted spectral model of GLEAM\,J024838--321336 for each epoch. The decreasing $\nu_\mathrm{p}$ from 2013 to September~2020 is consistent with an ejecta from the core traversing the jet.}
    \label{fig:024838_nu_p}
\end{figure}

GLEAM\,J024838--321336 showed variability during the 2020 monitoring unlike any other source; the SED for just the MWA frequency range is presented in Figure~\ref{fig:024838_mwa}. Most notable is the variability in the peak frequency, $\nu_p$, and flux density at the peak frequency, $S_{\nu_p}$. It appears $\nu_p$ shows a general trend of decreasing from 2013 right through to September~2020, shown in Figure~\ref{fig:024838_nu_p}. Additionally, $S_{\nu_p}$ increases until July~2020 and then is stable with the September~2020 SED. The odd behaviour of GLEAM\,J024838--321336 suggests a complex system or combination of mechanisms behind the variability. 

The first section of variability with an increasing $S_{\nu_p}$ and decreasing $\nu_p$ is consistent with an ejecta from the core cooling and expanding. Such an ejection would be emitting due to synchrotron radiation. Rearranging the equation for synchrotron emission, shown in Equation~\ref{eq:ssa_peakflux}, we can relate the energy of the emitting particles to the rest-frame brightness temperature of the emission. Therefore, as the temperature of the ejecta, $T_B$ decreases so too does the peak frequency according to: 
\begin{equation}
    k_B T_B \approx m_e c^2 \left(\frac{2\pi m_e\nu_p}{0.47eB\sin{\theta}} \right)^{1/2},
    \label{eq:Tb_SSA}
\end{equation}
\noindent
where $k_B$ is the Boltzmann constant and $m_e$, $e$ and $B\sin{\theta}$ were defined earlier by Equation~\ref{eq:ssa_peakflux} \citep{tingay2015spectral}.

Such a region slowly expanding and cooling would also be compact enough for ISS to be a dominant feature of the detected variability. The SEDs in July~2020 and September~2020 are fairly constant in shape with a decrease in $S_{\nu_p}$; this behaviour is consistent with RISS, as discussed in Section~\ref{sec:discussion_scintillators}. We suggest the variability of GLEAM\,J024838--321336 is due to both RISS and the cooling and expanding of a compact synchrotron-emitting region ejected from the core. Such a system would show a combination of increasing/decreasing flux density across the MWA band due to RISS with a slowly decreasing $\nu_p$. 

We note observations of X-ray binary systems, which can be considered analogous to AGN but on smaller scales, have detected ejecta from the core at multiple frequencies \citep{Fender2009XRBreview,tetarenko2019cygnus}. Lower frequencies detect emission further along the jets, away from the core. Monitoring of X-ray binary flares shows a lag in flares at lower frequencies consistent with the ejecta travelling along the jet. If the variability of GLEAM\,J024838--321336 is partly due to an ejection from the core slowly cooling and expanding, it is possible archival observations at higher frequencies ($\geq1$\,GHz) prior to the initial 2013 observations may have detected the initial ejection event from the core. Furthermore, follow-up high resolution imaging using VLBI would potentially be able to resolve such compact structures and test this interpretation.

\subsection{GLEAM J015445--232950}
\label{sec:discussion_j015445}
\begin{figure}
    \centering
    \includegraphics[width=\linewidth]{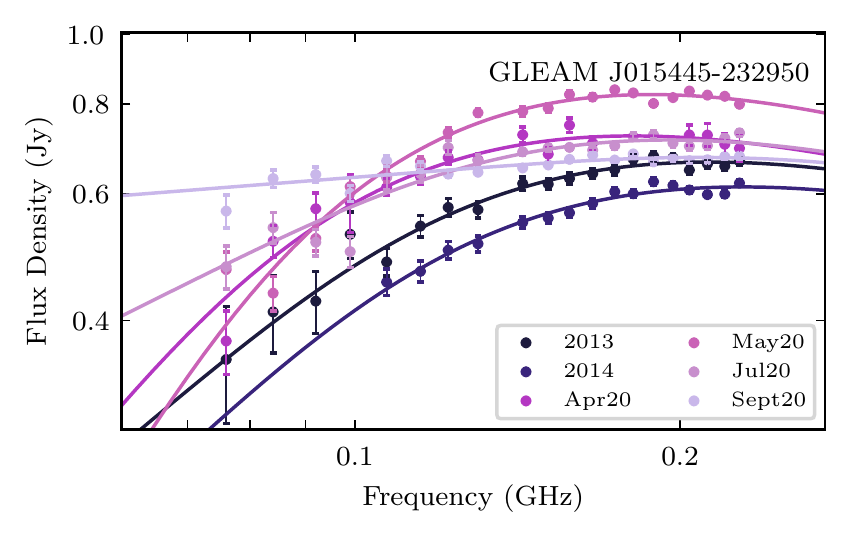}
    \caption{SED of only MWA flux densities for GLEAM\,J015445--232950. The constant spectral shape until May~2020 is consistent with interstellar scintillation. The changing spectral shape and spectral flattening in July and September~2020 is consistent with variations in the optical depth. }
    \label{fig:015445_mwa}
\end{figure}

Similar to \ffasource{}, GLEAM\,J015445--232950 shows two distinct forms of variability: a shift in flux density across the entire MWA spectra from 2013 to May~2020 (consistent with RISS), then an evolving spectral shape in July and September~2020. The SED of the MWA flux densities for GLEAM\,J015445--239250 are presented in Figure~\ref{fig:015445_mwa}. Interestingly, the spectral shape of GLEAM\,J015445--232950 in July and September appears to flatten rather than steepen like \ffasource{}. 

\begin{figure}
    \centering
    \includegraphics[width=\linewidth]{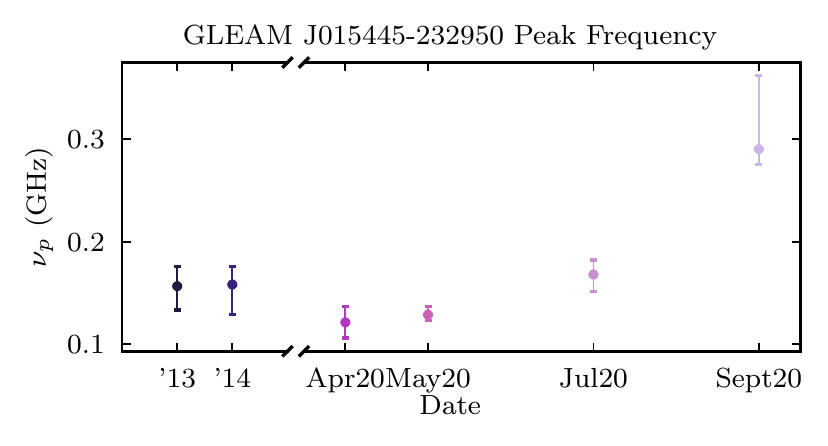}
    \caption{The measured peak frequency ($\nu_\mathrm{p}$) from the fitted spectral model of GLEAM\,J015445-232950 for each epoch.
    Similar to GLEAM\,J020507--110922, synchrotron emission can only explain the increasing $\nu_\mathrm{p}$, provided the source contract by $\leq10$\% (according to Equation~\ref{eq:ssa_peakflux}). }
    \label{fig:j015445_freqpeakvstime}
\end{figure}

Following the same logic described in Section~\ref{sec:discussion_j020507}, we consider changes in the synchrotron emission first. Figure~\ref{fig:j015445_freqpeakvstime} presents the variation of $\nu_p$ with time showing that the value of $\nu_p$ increased in September~2020 whilst $S_{\nu_p}$ decreases. This would require a significant \textit{decrease} in the size of the synchrotron emitting region, which is nonphysical. Furthermore, the changes in spectral shape would require improbably small-scale ($<25$\,mas) structures within the plasma for the variability to be due to scintillation. 

Lastly, we consider variations in the optical depth, $\tau_\mathrm{ff}$. Using Equation~\ref{eq:opacity_variation}, we calculate an upper limit for the opacity variation of 0.35 at 200\,MHz. While less than the opacity variation calculated for \ffasource{}, $\tau_\mathrm{ff}<0.35$ still suggests a significant gradient of varying optical depths in the absorbing ionized plasma. As noted in Section~\ref{sec:spectral_modelling}, $\tau_\mathrm{ff}$ is described by a power-law distribution with index $p$ and the spectral index in the optically thick regime $\alpha_\mathrm{thick}$, is proportional to $p$. A decrease in $\alpha_\mathrm{thick}$ is consistent with a decrease in $p$, or equivalently, a decrease in the optical depth. The spectral flattening of GLEAM\,J015445--232950 in July and September~2020 is consistent with a decrease in the optical depth suggesting GLEAM\,J015445--232950 is surrounded by an inhomogeneous free-free absorbing cloud. This is consistent with the results of spectral modelling, where GLEAM\,J015445-232950 is best explained by an inhomogeneous FFA model in five of the six epochs of MWA observations. The overall variability of GLEAM\,J015445--232950 can therefore be explained by a combination of RISS and a varying optical depth. 

\subsection{GLEAM J223933--451414}
\label{sec:discussion_j223933}
\begin{figure}
    \centering
    \includegraphics[width=\linewidth]{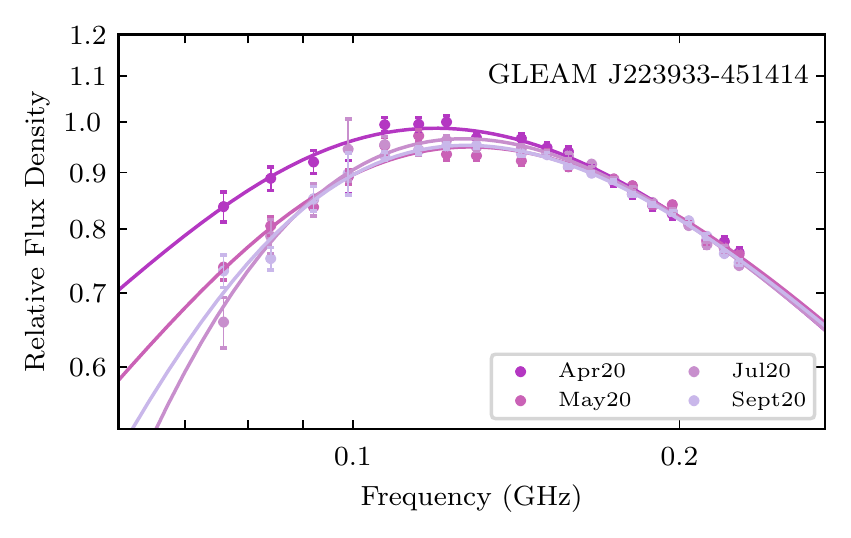}
    \caption{SED of only MWA flux densities and spectral models for GLEAM\,J223933--451414 in the 2020 epochs. May, July and September~2020 have had a constant factor added to the model and raw flux densities so each SED is similar in the optically thin region of the SED. The variability present below the spectral turnover despite the consistent flux densities in the optically-thin regime suggests some variability is due to changes in the absorption mechanism. }
    \label{fig:223933_mwa}
\end{figure}

GLEAM\,J223933--451414 showed variability consistent with sources discussed in Section~\ref{sec:discussion_scintillators}, as presented in Figure~\ref{fig:223933_sed}. The similar variability suggests it is varying due to RISS at MHz frequencies. However, there is also a notable steepening below the spectral turnover of ~130\,MHz, see Figure~\ref{fig:223933_mwa}. Similar to \ffasource{} and GLEAM\,J015445--232950, it is unlikely the variability below the turnover is due to synchrotron emission as the increased absorption would require decreases in the source size or an increase in the magnetic field. The larger modulation in the optically thick region suggests there are changes in the optical depth of GLEAM\,J223933--451414 due to an free-free absorbing medium. Using Equation~\ref{eq:opacity_variation}, we calculate a change in the free-free opacity of $\Delta\tau_\mathrm{ff}\lesssim0.45$ within the 2020 observations at 70\,MHz. Variability due to a changing free-free opacity suggests the physical mechanism producing the spectral turnover for GLEAM\,J223933--451414 is also due to free-free absorption. This interpretation is consistent with the spectral modelling. For each epoch, GLEAM\,J223933--451414 is best described by either a homogeneous or inhomogeneous free-free absorbing model with an exponential break. 

We therefore suggest that the variability of GLEAM\,J223933--451414 is due to two physical mechanisms: RISS, which produced a slow decrease in flux density across the MWA band during 2020; and changes in the optical depth due to an inhomogeneous free-free absorbing cloud surrounding GLEAM\,J223933--451414. High resolution images on mas scales could search for the presence of a feature compact enough for variability due to ISS. Previous detections of dust surrounding AGN have been made via observations of absorption features in the infrared spectra \citep{2021MNRAS.502.2508Z,2015P&SS..116...97M}. Testing for prominent HI gas and other absorption features would determine whether the variability is caused by variations in the optical depth. Furthermore, we recommend continued monitoring of GLEAM\,J223933--451414, particularly below the spectral turnover, to detect and characterise any variability in the absorption that could be attributed to variations in the optical depth. 

\section{Conclusions}
\label{sec:conclusion}
R21 identified variable sources using two epochs of MWA observations separated by approximately one year. Subsequently, we have monitored 15 PS sources during 2020 with the MWA and the ATCA to search for and characterise spectral variability across 72\,MHz -- 10\,GHz. We found 13 of the 15 targets continued to show variability at MHz frequencies. We detect no significant variability at GHz frequencies for any source on timescales of weeks to months. 

We discussed the nature of ISS and the spectral variations it can produce. We determine it is unlikely to create changes in spectral shape, particularly on month long timescales at MHz frequencies unless there are small structures within the ISM on scales of $\sim$AU. We find nine sources show slow trends of either increasing or decreasing flux densities across the entire MWA bandwidth with a constant spectral shape. Slow variable trends at 150\,MHz over the course of $\sim$1\,year is consistent with a compact feature approximately 25\,mas in size scintillating due to ISS. We therefore attribute this variability entirely to ISS. To confirm, we detect intra-day variability of GLEAM\,J001513--472706 at 5 and 9\,GHz with the ATCA, also consistent with ISS. The short snapshot observations of targets in the 2020 monitoring meant there was insufficient sampling for searching for hourly variability in these epochs. 

We discuss GLEAM\,J020507--110922 in detail due to the sudden change in spectral shape in September~2020 and the increase in peak frequency by $\sim$100\,MHz. We consider variability due to changes in the synchrotron emission/absorption, ISS, an ESE and variations in the optical depth. We determine two likely origins for the variability of GLEAM\,J020507--110922: ISS and changes in the optical depth. The variability of GLEAM\,J020507--110922 prior to September~2020 is consistent with ISS, however the change in spectral shape from July to September suggests either small structures within the scintillating screen comparable to the structures that would produce an ESE. The lack of intra-day variability at 5 and 9\,GHz and the increase in peak frequency supports the conclusion of a second origin of variability. The change in spectral shape of GLEAM\,J020507--110922 from July to September~2020 is consistent with a varying optical depth due to an inhomogeneous free-free absorbing cloud, where a `clump' of either higher electron density or cooler electron temperature has moved into the line of sight. We conclude the origins of spectral variability for GLEAM\,J020507--110922 are due to both ISS and an inhomogeneous ionized cloud surrounding the source. We combine the evidence of the most likely spectral model, an inhomogeneous FFA model with an exponential break, with the origins of spectral variability to determine the cause of the spectral turnover as an inhomogeneous FFA model. 

We find GLEAM\,J015445-232950 and GLEAM\,J223933--451414 show similar variability to GLEAM\,J020507--110922. GLEAM\,J015445--232950 shows variability consistent with ISS until July~2020 then a flattening of the spectral shape below the spectral turnover in July and September~2020. As with GLEAM\,J020507--110922, we conclude the origins of the spectral variability is most likely due to a combination of ISS and variations in the optical depth from an inhomogeneous free-free absorbing cloud. However, as the absorption decreases, it is likely either a `clump' of hotter temperature electrons  with a lower electron density has moved into the line of sight. Similarly, GLEAM\,J223933--451414 shows a constant spectral shape above the turnover frequency but a steepening below the spectral turnover. We conclude, both GLEAM\,J015445--232950 and GLEAM\,J223933--4511414 are best explained by an inhomogeneous FFA spectral model with an exponential break based on their spectral fitting and spectral variability. 

We investigate the variable peak frequency of GLEAM\,J024838--321336. The decreasing peak frequency is consistent with a cooling ejecta travelling along the jet, which is also compact enough to scintillate due to ISS. Due to the likely origins of the spectral variability and the spectral fitting finding an SSA model as the most likely, we determine the most likely explanation for the absorption of GLEAM\,J024838--321336 is due to synchrotron self absorption. 

The results of this variability study show the large spectral coverage, particularly at MHz frequencies, is key to determining the origins of the variability. Furthermore, PS sources continue to be a rich source of variability, particularly showing distinct forms of variability in the optically thick and thin regimes. We show that combining spectral modelling with spectral variability is a novel and powerful tool to determine the likely cause of absorption of PS sources. We recommend future observations of spectral variability of PS sources, particularly in the optically thick regime, to determine the absorption mechanism. 

In the SKA era, as large-scale surveys become feasible, it is crucial we design surveys with large spectral and temporal coverage in order to adequately sample spectral variability. In particular, we should design surveys with cadences that probe timescales relating to specific types of variability paired with complementary spectral coverage. In particular, for scintillation monitoring on six monthly to yearly cadences at megahertz frequencies compared to hour to day cadences at gigahertz frequencies. Likewise, monitoring on monthly cadences at megahertz frequencies for variability due to free-free absorption. This paper highlights the value of low (MHz) frequency spectral coverage over month-year-decade long timescales with high (GHz) frequency observations on minutes-hours-days (ideally simultaneously) in distinguishing the origins of variability.

\section*{Acknowledgements}
KR thanks Paul Hancock and John Morgan for discussions on scintillation and comments on this paper. 
KR acknowledges a Doctoral Scholarship and an Australian Government Research Training Programme scholarship administered through Curtin University of Western Australia. JRC thanks the Nederlandse Organisatie voor Wetenschappelijk Onderzoek (NWO) for support via the Talent Programme Veni grant. NHW is supported by an Australian Research Council Future Fellowship (project number FT190100231) funded by the Australian Government. This scientific work makes use of the Murchison Radio-astronomy Observatory, operated by CSIRO. We acknowledge the Wajarri Yamatji people as the traditional owners of the Observatory site. Support for the operation of the MWA is provided by the Australian Government (NCRIS), under a contract to Curtin University administered by Astronomy Australia Limited. We acknowledge the Pawsey Supercomputing Centre, which is supported by the Western Australian and Australian Governments. This research made use of NASA's Astrophysics Data System, the VizieR catalog access tool, CDS, Strasbourg, France. We also make use of the \textsc{IPython} package \citep{Ipythoncite}; SciPy \citep{2020SciPy-NMeth};  \textsc{Matplotlib}, a \textsc{Python} library for publication quality graphics \citep{Hunter:2007}; \textsc{Astropy}, a community-developed core \textsc{Python} package for astronomy \citep{astropy:2013, astropy:2018}; \textsc{pandas}, a data analysis and manipulation \textsc{Python} module \citep{reback2020pandas,mckinney-proc-scipy-2010}; and \textsc{NumPy} \citep{vaderwalt_numpy_2011}. We also made extensive use of the visualisation and analysis packages DS9\footnote{\href{ds9.si.edu}{http://ds9.si.edu/site/Home.html}} and Topcat \citep{2005ASPC..347...29T}. This work was compiled in the useful online \LaTeX{} editor Overleaf.

\section*{Data Availability}
The raw data underlying this article are available in the ATCA Online Archive (ATOA) for project code C3333 and the MWA All-Sky Virtual Observatory (ASVO) for project code G0067. Derived and processed data products, including fits images and measured flux densities, are available upon request.



\bibliographystyle{mnras}
\bibliography{bibliography} 




\appendix
\section{Supplementary Spectral Energy Distributions (SEDs)}
\label{sec:app:seds}
The SEDs for the 15 sources nearby ($\leq1$\,degree) the targets of interest can be found in the supplementary online materials. Nearby sources were used to confirm that the variability observed was unique to the source and not due to the data processing. Several nearby sources were inspected and these 15 nearby sources are included as examples. Uncertainties are calculated from the flux density uncertainties estimated from the Background And Noise Estimation Tool (\textsc{BANE}) plus a $2\%$ flux density error added in quadrature. The $2\%$ flux density measurement was used to account for the systematic and random noise of the images across the MWA band and is the internal uncertainty on the GLEAM flux density measurements. 

\bsp	
\label{lastpage}
\end{document}